\documentclass[aps,prl,twocolumn,superscriptaddress,nofootinbib,floatfix]{revtex4-2}

\usepackage{amsmath,amssymb,bm}
\usepackage[colorlinks=true,linkcolor=blue,citecolor=blue,urlcolor=blue]{hyperref}

\newcommand{\Ft}{\widetilde F}
\newcommand{\Dc}{\mathcal{D}_\chi}
\newcommand{\vp}{\varpi}
\newcommand{\Sn}{S^{\mu\nu}_{(n)}}
\newcommand{\dpr}{\delta'(p^2)}

\begin{document}

\title{Equilibrium relativistic spin magnetohydrodynamics from chiral kinetic theory}

\author{Gopal Mistry}
\email{gopal.mistry@niser.ac.in}
\affiliation{School of Physical Sciences, National Institute of Science Education and Research, An OCC of Homi Bhabha National Institute, Jatni-752050, India}

\author{Amaresh Jaiswal}
\email{a.jaiswal@niser.ac.in}
\affiliation{School of Physical Sciences, National Institute of Science Education and Research, An OCC of Homi Bhabha National Institute, Jatni-752050, India}

\date{\today}

\begin{abstract}
We develop a microscopic and thermodynamically consistent framework for relativistic equilibrium spin magnetohydrodynamics of massless chiral fermions, directly from first-order chiral kinetic theory. Starting from the covariant Wigner function, we recast the chiral kinetic equation as a pure streaming equation on a magnetically shifted mass shell, incorporating both the magnetic-moment energy and the Berry-curvature phase-space measure. This formulation yields a kinetic $H$-theorem and an entropy current with vanishing divergence in global equilibrium. We construct the equilibrium distribution for an arbitrary spin frame and derive the associated currents, energy--momentum tensor, spin tensor, entropy current, and frame-independent spin polarization, including the magnetic splitting between particles and antiparticles. Thermodynamic consistency further determines the second-order magnetovortical responses, revealing a relativistic Barnett magnetization, a temperature- and chemical-potential-independent anomalous charge density, and rotation-induced density corrections that reproduce exact field-theory results. Our framework provides a unified microscopic foundation for relativistic equilibrium spin magnetohydrodynamics and establishes a direct link between quantum kinetic theory, spin polarization, magnetization, and anomalous transport.
\end{abstract}

\maketitle

%%%%%%%%%%%%%%%%%%%%%%%%%%%%%%%%%%%%%%%%%%%%%%%%%%

\textit{Introduction:}
Quantum anomalies provide one of the deepest connections between microscopic symmetries and macroscopic transport in quantum many-body systems, giving rise to nondissipative transport phenomena that couple spin, vorticity, and electromagnetic fields. In chiral matter they induce currents along a magnetic field and along the vorticity, i.e., the chiral magnetic (CME)~\cite{Vilenkin:1980fu, Fukushima:2008xe} and chiral vortical (CVE)~\cite{Vilenkin:1979ui, Son:2009tf, Landsteiner:2011cp} effects. These anomaly-induced responses are expected to play a central role in the dynamics of the quark--gluon plasma produced in relativistic heavy-ion collisions~\cite{Kharzeev:2015znc}, Weyl and Dirac semimetals~\cite{Son:2012bg, Landsteiner:2013sja, Sekine:2017eeo, Armitage:2017cjs, bernevig2013topological}, rapidly rotating neutron stars~\cite{Kaminski:2014jda}, and the early Universe~\cite{Joyce:1997uy, Boyarsky:2011uy}. The observation of global $\Lambda$ polarization~\cite{STAR:2017ckg} has further motivated relativistic spin hydrodynamics~\cite{Florkowski:2017ruc, Florkowski:2017dyn, Florkowski:2018ahw, Florkowski:2018fap, Hattori:2019lfp, Li:2020eon, Singh:2022ltu} and its magnetohydrodynamic extension~\cite{Bhadury:2022ulr, Buzzegoli:2022qrr}, in which spin polarization, magnetization, and anomalous transport must be described within one framework. A microscopic understanding of how spin polarization, magnetization, and anomalous transport emerge within relativistic hydrodynamics is therefore a fundamental problem at the interface of quantum field theory, kinetic theory, and nonequilibrium many-body physics.

Chiral kinetic theory (CKT) provides a microscopic framework for such a description~\cite{Stephanov:2012ki, Son:2012wh, Son:2012zy, Chen:2012ca, Gao:2012ix, Chen:2014cla, Chen:2015gta, Hidaka:2016yjf, Huang:2018wdl, Gao:2020pfu, Hidaka:2022dmn, Yang:2024hmd}. At first order in $\hbar$, quantum effects manifest through the side-jump contribution to the current, the magnetic-moment correction to the quasiparticle energy, and the Berry-curvature modification of the phase-space measure. CKT has also been used to derive spin hydrodynamics~\cite{Shi:2020htn}. A thermodynamically consistent macroscopic theory, however, requires an equilibrium solution and a kinetic equation compatible with the second law of thermodynamics. In electromagnetic fields, existing covariant formulations of the chiral kinetic equation (CKE) involve $\delta'(p^2)$ terms or equivalent momentum derivatives that obscure the Liouville structure of phase-space evolution and complicate the formulation of the kinetic $H$-theorem. Thus, a unified framework for relativistic equilibrium spin magnetohydrodynamics, derived directly from first-order CKT and consistent with thermodynamics, remains lacking for chiral fermions.

In this Letter, we derive a thermodynamically consistent framework for relativistic equilibrium spin magnetohydrodynamics of massless chiral fermions in $(3+1)$ dimensions, directly from first-order chiral kinetic theory. Starting from the covariant Wigner function~\cite{Huang:2018wdl, Shi:2020htn, Yang:2024hmd}, we recast the chiral kinetic equation as a pure streaming equation on a magnetically shifted mass shell, incorporating both magnetic-moment and Berry-curvature effects. This formulation elucidates the underlying Liouville structure and yields a kinetic $H$-theorem, with an entropy current whose divergence vanishes in global equilibrium~\cite{Jaiswal:2013fc}. For an arbitrary spin frame, we construct the equilibrium distribution and derive the associated constitutive relations, including the vector and axial currents, energy--momentum and spin tensors, and frame-independent fermion polarization with a magnetic particle--antiparticle splitting. 

The framework consistently reproduces the chiral magnetic and vortical effects and determines the second-order magnetovortical responses, including relativistic Barnett magnetization~\cite{Barnett:1915}, anomalous charge redistribution, and rotation-induced density corrections that reproduce exact field-theory results. We demonstrate that the field-induced magnetization produces a Pauli-type paramagnetic contribution from moment polarization, rather than exhibiting a Curie law. By establishing a link between covariant quantum kinetic theory, spin polarization, and magnetovortical transport, our work provides a unified microscopic foundation for relativistic equilibrium spin magnetohydrodynamics, with applications to heavy-ion collisions, astrophysics, and condensed-matter systems.

Throughout this work, we employ natural units with $c=k_B=1$ while keeping $\hbar$ explicit as the semiclassical counting parameter. We adopt the metric convention $g^{\mu\nu}=\mathrm{diag}(1,-1,-1,-1)$, and normalize the totally anti-symmetric Levi--Civita tensor as $\epsilon^{0123}=-\epsilon_{0123}=1$. Three-vectors are denoted in boldface, and the dot indicates the scalar product for both three- and four-vectors, i.e., $a\cdot b \equiv a^0b^0-\mathbf{a}\cdot\mathbf{b}$. The dual of the electromagnetic field tensor is defined as $\Ft^{\mu\nu}=\frac12\epsilon^{\mu\nu\alpha\beta}F_{\alpha\beta}$.

%%%%%%%%%%%%%%%%%%%%%%%%%%%%%%%%%%%%%%%%%%%%%%%%%%

\medskip

\textit{Wigner function and chiral currents:}
For massless fermions of charge $Q$, the gauge-covariant Wigner function obeys~\cite{Vasak:1987um, Elze:1986hq, Elze:1986qd, Zhuang:1995pd, Florkowski:1995ei}
\begin{equation}
\gamma_\mu\Big(p^\mu+\frac{i\hbar}{2}\nabla^\mu\Big)W(x,p)=0,\quad~~
\nabla_\mu\equiv\partial_\mu-QF_{\mu\nu}\partial_p^{\nu},
\label{eq:wigner}
\end{equation}
where $\partial_p^\nu\equiv\partial/\partial p_\nu$ and quantum corrections to $\nabla_\mu$ are of order $\hbar^2\partial^2F$. Expanding in the Clifford basis~\cite{Vasak:1987um, Weickgenannt:2019dks},
\begin{equation}
W=\frac14\left[\mathcal F+i\gamma^5\mathcal P+\gamma^\mu V_\mu+\gamma^5\gamma^\mu A_\mu+\frac12\sigma^{\mu\nu}S_{\mu\nu}\right], 
\end{equation}
the chiral currents $J^\mu_\chi=\frac12(V^\mu+\chi A^\mu)$, with $\chi=\pm1$, decouple from other components and satisfy~\cite{Gao:2012ix, Huang:2018wdl, Gao:2020pfu}
\begin{align}
&p_\mu J^\mu_\chi=0,\qquad \nabla_\mu J^\mu_\chi=0,\nonumber\\
&p^\mu J^\nu_\chi-p^\nu J^\mu_\chi=-\frac{\chi\hbar}{2}\epsilon^{\mu\nu\rho\sigma}\nabla_\rho J_{\chi\sigma}.
\label{eq:Jeqs}
\end{align}

Solving Eqs.~\eqref{eq:Jeqs} to first order in $\hbar$ gives [Supplemental Material (SM), Sec.~S2]~\cite{Hidaka:2016yjf, Huang:2018wdl}
\begin{equation}
J^\mu_\chi=\delta(p^2)\big[p^\mu+\chi\hbar\,\Sn\nabla_\nu\big]f_\chi+\chi\hbar Q\,\Ft^{\mu\nu}p_\nu\,\dpr f_\chi ,
\label{eq:J}
\end{equation}
where $\Sn=\epsilon^{\mu\nu\rho\sigma}p_\rho n_\sigma/(2\,n\!\cdot\! p)$ is the spin tensor of a massless fermion defined with respect to a timelike frame vector $n^\mu$~\cite{Costa:2017kdr}. In the above equation for chiral current, the second term represents the side-jump contribution, while the last term accounts for the magnetic-moment correction. Although the distribution function $f_\chi$ depends on the choice of spin-frame vector $n^\mu$, the physical current $J^\mu_\chi$ is frame independent. In particular, under a change of frame $n\to n'$, the current remains invariant provided the distribution transforms as $f_{\chi(n')}=f_{\chi(n)}+\chi\hbar\,\epsilon^{\alpha\beta\gamma\delta}p_\alpha n_\beta n'_\gamma\nabla_\delta f_\chi/[2(n\cdot p)(n'\cdot p)]$.

%%%%%%%%%%%%%%%%%%%%%%%%%%%%%%%%%%%%%%%%%%%%%%%%%%

\medskip

\textit{Kinetic equation in streaming form:}
Inserting Eq.~\eqref{eq:J} into $\nabla_\mu J^\mu_\chi=0$ and using the Bianchi and Schouten identities, all terms proportional to $\delta'(p^2)$ either cancel or combine with $\delta(p^2)$ into a shifted mass shell (SM, Sec.~S3). The result holds for an arbitrary phase-space function $f_\chi$, leading to
\begin{equation}
\nabla_\mu J^\mu_\chi=\delta\big(p^2-\chi\hbar Q F_{\alpha\beta}S^{\alpha\beta}_{(n)}\big)\,\Dc f_\chi+\mathcal{O}(\hbar^2),
\label{eq:divJ}
\end{equation}
with the streaming operator given by
\begin{align}
\Dc=p^\mu\nabla_\mu&+\chi\hbar\Big[\partial_\mu\Sn+\frac{Q}{n\cdot p}E_\mu\Sn\Big]\nabla_\nu \nonumber\\
&-\chi\hbar Q\,\Sn(\partial_\mu F_{\nu\lambda})\,\partial_p^\lambda ,
\label{eq:D}
\end{align}
where $E_\mu\equiv F_{\mu\nu}n^\nu$. The chiral kinetic equation can therefore be written as
\begin{equation}
\Dc f_\chi=0\quad\text{on}\quad p^2=\chi\hbar Q\,\frac{B_{(n)}\!\cdot p}{n\cdot p},
\label{eq:CKE}
\end{equation}
with $B^\mu_{(n)}\equiv\Ft^{\mu\nu}n_\nu$, since $F_{\alpha\beta}S^{\alpha\beta}_{(n)}=\left(B_{(n)}\!\cdot\! p\right)/(n\!\cdot\! p)$. The form of the chiral kinetic equation derived above generalizes the results of Ref.~\cite{Huang:2018wdl}.

Note that two features of Eqs.~\eqref{eq:divJ}--\eqref{eq:CKE} are particularly important. First, $\Dc$ is a first-order differential operator without any multiplicative term such that it satisfies the chain rule, $\Dc\Phi(f)=\Phi'(f)\,\Dc f$ for any function $\Phi$. Second, the magnetic-moment correction to the quasiparticle energy and the Berry-curvature modification of the phase-space measure are encoded in the magnetically shifted mass shell. To illustrate this, consider constant electromagnetic fields and a constant spin-frame vector $n^\mu$ with $E_\mu=0$. In the rest frame of $n^\mu$, the shifted shell yields the quasiparticle energy $E_{\bm p}=|\bm p|(1-Q\bm B\cdot\bm\Omega_\chi)$, where $\bm\Omega_\chi=\chi\hbar\hat{\bm p}/(2|\bm p|^2)$ is the Berry curvature. In this limit, Eq.~\eqref{eq:CKE} reduces to the familiar three-dimensional chiral kinetic equation (SM, Sec.~S3),
\begin{equation}
\big(\partial_t+\dot{\bm x}\cdot\bm\nabla_{\bm x}+\dot{\bm p}\cdot\bm\nabla_{\bm p}\big)f_\chi=0,
\label{eq:CKE3}
\end{equation}
with $\sqrt\gamma\,\dot{\bm x}=\tilde{\bm v}+Q(\tilde{\bm v}\cdot\bm\Omega_\chi)\bm B$, $\sqrt\gamma\,\dot{\bm p}=Q\tilde{\bm v}\times\bm B$, $\tilde{\bm v}=\partial E_{\bm p}/\partial\bm p$, and $\sqrt\gamma=1+Q\bm B\cdot\bm\Omega_\chi$. These are the standard chiral kinetic equations~\cite{Stephanov:2012ki, Son:2012wh, Son:2012zy}. The associated phase-space flow obeys Liouville's theorem given by $\partial_t\sqrt\gamma+\bm\nabla_{\bm x}\cdot(\sqrt\gamma\dot{\bm x})+\bm\nabla_{\bm p}\cdot(\sqrt\gamma\dot{\bm p})=0$, which ensures the equivalence of the streaming and continuity forms of the kinetic equation.

%%%%%%%%%%%%%%%%%%%%%%%%%%%%%%%%%%%%%%%%%%%%%%%%%%

\medskip

\textit{Entropy current:}
In order to analyze the entropy production, we begin by noting that $\nabla_\mu J^\mu_\chi\left[\Phi(f_\chi)\right]$ in Eq.~\eqref{eq:divJ} holds for any argument $\Phi(f_\chi)$. Accordingly, we define the entropy current using the same functional,
\begin{equation}
\mathcal H^\mu=-\sum_{\chi}\int_P\Big(J^\mu_\chi\left[\Phi(f_\chi)\right]+\bar J^\mu_\chi\left[\Phi(\bar f_\chi)\right]\Big),
\label{eq:Smu}
\end{equation}
where $\Phi(f)=f\ln f+(1-f)\ln(1-f)$ is the Fermi–Dirac entropy function. The positive-energy momentum integral is defined as $\int_P\equiv2\int d^4p\,\theta(p^0)/(2\pi)^3$ and the antiparticle current $\bar J^\mu_\chi$ is obtained from $J^\mu_\chi$ by setting $(Q,\chi)\to(-Q,-\chi)$ with $\bar f_\chi$ denoting the corresponding antiparticle distribution. 

For $F_{\mu\nu}\Ft^{\mu\nu}=0$, the momentum-space total-derivative contribution arising from $\partial_\mu = \nabla_\mu + QF_{\mu\lambda} \partial_p^\lambda$ vanishes upon integration. Consequently, applying Eq.~\eqref{eq:divJ} together with the chain rule for $\Dc$ yields
\begin{align}
\partial_\mu\mathcal H^\mu=\sum_\chi\int_P&\delta\!\left(p^2-\chi\hbar QF_{\alpha\beta}S^{\alpha\beta}_{(n)}\right) \ln\!\left[\frac{1-f_\chi}{f_\chi}\right]\Dc f_\chi\nonumber\\
&+(\text{antiparticles}).
\label{eq:dS}
\end{align}
For collisionless evolution, we see that $\partial_\mu\mathcal H^\mu=0$ up to ${\cal O}(\hbar^2)$. With a collision kernel, $\Dc f_\chi=C_\chi[f]$, Eq.~\eqref{eq:dS} becomes the usual $H$-theorem, i.e., the entropy production is the moment of $C_\chi$ weighted by $\ln[(1-f)/f]$, and it vanishes if and only if this logarithm is a linear combination of summational invariants. This is precisely the equilibrium constructed below.

%%%%%%%%%%%%%%%%%%%%%%%%%%%%%%%%%%%%%%%%%%%%%%%%%%

\medskip

\textit{Global equilibrium:}
We seek an equilibrium distribution of the Fermi–Dirac form, $f_\chi=[e^{g_\chi}+1]^{-1}$, where $g_\chi$ is constructed from the collision invariants associated with charge, four-momentum, and total angular momentum (including both orbital and spin contributions). Using the chain rule, $\Dc f=f'(g)\,\Dc g$, the kinetic equation \eqref{eq:CKE} then requires $\Dc g_\chi=0$. To first order in $\hbar$ the solution is (SM, Sec.~S5)
\begin{equation}
g_\chi=\beta_\mu p^\mu-\alpha_\chi-\frac{\chi\hbar}{2}\,\vp_{\mu\nu}S^{\mu\nu}_{(n)},
\label{eq:geq}
\end{equation}
where $\vp_{\mu\nu}=-\frac12(\partial_\mu\beta_\nu-\partial_\nu\beta_\mu)$ is the thermal vorticity. Global equilibrium further requires the constraints
\begin{equation}
\partial_\mu\beta_\nu+\partial_\nu\beta_\mu=0,\quad
\partial_\mu\alpha_\chi=-QF_{\mu\nu}\beta^\nu,\quad
\pounds_\beta F_{\mu\nu}=0 ,
\label{eq:Killing}
\end{equation}
where $\beta^\mu$ is a Killing vector and $\vp_{\mu\nu}$ is constant in flat spacetime. The second condition describes the equilibrium balance between the gradient of the chemical potential and the electromagnetic force, while the last requires the electromagnetic field to be invariant under the flow generated by $\beta^\mu$ [SM, Eq.~(S33)]. For a rotating fluid, a magnetic field aligned with the rotation axis provides an example of a configuration compatible with this symmetry. The antiparticle distribution follows from the replacements $(\alpha_\chi,\chi)\to(-\alpha_\chi,-\chi)$. 

Writing $\beta_\mu=b_\mu+\vp_{\mu\nu}x^\nu$, we obtain 
\begin{equation}
g_\chi=b\cdot p-\alpha_\chi-\frac12\vp_{\mu\nu} \left(x^\mu p^\nu-x^\nu p^\mu+\chi\hbar S^{\mu\nu}_{(n)}\right).
\end{equation}
Therefore, we see that $\vp_{\mu\nu}$ naturally couples to the total angular momentum. The spin-dependent term ensures that $f_\chi$ transforms under $n\to n'$ according to the required side-jump transformation, leaving the physical equilibrium state frame independent. Although no explicit electromagnetic coupling appears in $g_\chi$, the spin--magnetic interaction is encoded in the magnetically shifted mass shell of Eq.~\eqref{eq:CKE}. In the local rest frame, we identify $\beta^\mu=u^\mu/T$, $\alpha_\chi=\mu_\chi/T$ and choose the spin frame $n^\mu=u^\mu$. Further, we use the definitions $\omega^\mu=-\frac T2\epsilon^{\mu\nu\rho\sigma}u_\nu\vp_{\rho\sigma}=\frac12\epsilon^{\mu\nu\rho\sigma}u_\nu\partial_\rho u_\sigma$ and $B^\mu=\Ft^{\mu\nu}u_\nu$. The resulting on-shell distribution becomes
\begin{equation}
f_\chi=\left\{\exp\!\left[\frac{|\bm p|-\mu_\chi}{T}-\frac{\chi\hbar}{2T}\,\hat{\bm p}\cdot\!\left(\bm\omega+\frac{Q\bm B}{|\bm p|}\right)\right]+1\right\}^{-1}\!,
\label{eq:frest}
\end{equation}
where $\hat{\bm p}=\bm p/|\bm p|$. The corresponding momentum-space integrals are weighted by the Berry-curvature-modified measure $\sqrt\gamma$. The above expression makes explicit the coupling of the fermion spin, $\chi\hbar\hat{\bm p}/2$, to vorticity and to the magnetic field, with the latter governed by the effective magnetic coupling $Q/|\bm p|$. The on-shell distribution in Eq.~\eqref{eq:frest} can be written in covariant form as
\begin{equation}
f_\chi=\dfrac{1}{\exp\!\left[\dfrac{u\!\cdot\! p-\mu_\chi}{T}+\dfrac{\chi\hbar}{2T}\!\left(\dfrac{\omega\!\cdot\! p}{u\!\cdot\! p}+Q\dfrac{B\!\cdot\! p}{(u\!\cdot\! p)^2}\right)\right]+1},
\end{equation}
with the measure $\sqrt\gamma$.

%%%%%%%%%%%%%%%%%%%%%%%%%%%%%%%%%%%%%%%%%%%%%%%%%%

\medskip

\textit{Constitutive relations:}
Macroscopic quantities are defined via the moments 
\begin{align}
j^\mu_\chi &= \int_P \left(J^\mu_\chi[f_\chi] - \bar J^\mu_\chi[\bar f_\chi] \right), \\
T^{\mu\nu} &= \sum_\chi\int_P p^\nu \left(J^\mu_\chi[f_\chi] + \bar J^\mu_\chi[\bar f_\chi] \right).
\end{align}
We evaluate the moments at a point where the comoving electric field $E^\mu\equiv F^{\mu\nu}u_\nu$ vanishes, as appropriate in ideal magnetohydrodynamics. The resulting momentum integrals reduce to standard Fermi integrals, whose particle–antiparticle combinations yield exact polynomials in $\mu_\chi/T$. Introducing the vector and axial chemical potentials, $\mu=(\mu_++\mu_-)/2$ and $\mu_5=(\mu_+-\mu_-)/2$, the vector and axial currents, defined by $j^\mu=\sum_\chi j^\mu_\chi$ and $j^\mu_5=\sum_\chi\chi j^\mu_\chi$, are
\begin{align}
j^\mu&=n\,u^\mu+\hbar\xi\,\omega^\mu+\hbar\xi_B B^\mu, \label{eq:currents1}\\
j_5^\mu&=n_5u^\mu+\hbar\xi_5\,\omega^\mu+\hbar\xi_{B5} B^\mu,
\label{eq:currents}
\end{align}
with
\begin{align}
n&=\frac{\mu^3+3\mu\mu_5^2}{3\pi^2}+\frac{\mu T^2}{3},& n_5&=\frac{\mu_5^3+3\mu^2\mu_5}{3\pi^2}+\frac{\mu_5T^2}{3},\nonumber\\
\xi&=\frac{\mu\mu_5}{\pi^2},& \xi_5&=\frac{\mu^2+\mu_5^2}{2\pi^2}+\frac{T^2}{6},\nonumber\\
\xi_B&=\frac{Q\mu_5}{2\pi^2},& \xi_{B5}&=\frac{Q\mu}{2\pi^2}.
\label{eq:coeffs}
\end{align}
These are the CVE, CME and chiral separation effect (CSE) coefficients of free Weyl fermions~\cite{Vilenkin:1979ui, Vilenkin:1980fu, Fukushima:2008xe, Son:2009tf, Landsteiner:2011cp}. The vortical response receives one-third of its contribution from the spin--vorticity coupling in the equilibrium distribution and the remaining two-thirds from the side-jump current. On the other hand, the magnetic response originates entirely from the $\delta'(p^2)$ contribution in Eq.~\eqref{eq:J}. The coefficients satisfy the thermodynamic relations $\xi=\frac12\partial n_5/\partial\mu$, $\xi_5=\frac12\partial n/\partial\mu$, $\xi_B=\frac Q2\partial\xi/\partial\mu$, and $\xi_{B5}=\frac Q2\partial\xi_5/\partial\mu$. The first-order response is therefore completely determined by the zeroth-order equation of state.

The energy-momentum tensor is obtained as
\begin{align}
T^{\mu\nu}=\;&\varepsilon u^\mu u^\nu-P\Delta^{\mu\nu}+\hbar\big(q^\mu u^\nu+u^\mu q^\nu\big)\nonumber\\
&+\frac{\hbar n_5}{2}\big(\omega^\mu u^\nu-u^\mu\omega^\nu-\epsilon^{\mu\nu\alpha\beta}u_\alpha a_\beta\big),
\label{eq:Tmunu}
\end{align}
where $\Delta^{\mu\nu}=g^{\mu\nu}-u^\mu u^\nu$, $a^\mu=u^\nu\partial_\nu u^\mu=T\vp^{\mu\nu}u_\nu$, $P=\varepsilon/3$, with
\begin{equation}
\varepsilon=\frac{7\pi^2T^4}{60}+\frac{T^2(\mu^2+\mu_5^2)}{2}+\frac{\mu^4+6\mu^2\mu_5^2+\mu_5^4}{4\pi^2}.
\label{eq:eps}
\end{equation}
The energy flow vector is given by $q^\mu=n_5\omega^\mu+\mu\,\xi_B B^\mu$. For a single Weyl fermion of chirality $\chi$, the magnetic contribution to the energy flow is $\chi Q(\mu_\chi^2/8\pi^2+T^2/24)B^\mu$. The $T^2$ contribution is associated with the mixed axial--gravitational anomaly~\cite{Landsteiner:2011cp}, which also plays an important role in anomalous thermoelectric transport in Weyl semimetals~\cite{Gooth:2017mbd}. 

The antisymmetric part of the energy--momentum tensor is related to the spin tensor,
\begin{equation}
S^{\lambda\mu\nu}=-\frac{\hbar}{2}\epsilon^{\lambda\mu\nu\sigma}j_{5\sigma},
\label{eq:spin}
\end{equation}
through the local angular-momentum balance relation $\partial_\lambda S^{\lambda\mu\nu}=T^{\nu\mu}-T^{\mu\nu}$, i.e., the total angular momentum is conserved (SM, Sec.~S8). Finally, the entropy current defined in Eq.~\eqref{eq:Smu} evaluates to
\begin{equation}
\mathcal H^\mu=s\,u^\mu+\frac{\hbar\,\mu_5T}{3}\,\omega^\mu,\qquad s=\frac{\varepsilon+P-\mu n-\mu_5n_5}{T}.
\label{eq:entropy}
\end{equation}
Note that the equilibrium entropy current contains no magnetic contribution, and its divergence vanishes, in accordance with the kinetic entropy relation in Eq.~\eqref{eq:dS}.

%%%%%%%%%%%%%%%%%%%%%%%%%%%%%%%%%%%%%%%%%%%%%%%%%%

\medskip

\textit{Spin polarization:}
The axial component $A^\mu=\sum_\chi\chi J^\mu_\chi$ is the phase-space density of spin; with its antiparticle counterpart $\bar A^\mu=\sum_\chi\chi\bar J^\mu_\chi$, the spin density is $s^\mu=\frac\hbar2\int_P(A^\mu-\bar A^\mu)=\frac\hbar2j^\mu_5$. In equilibrium the frame dependence of $f_\chi$ cancels against that of the side-jump current, and we obtain (SM, Sec.~S9)
\begin{align}
A^\mu&=p^\mu\delta(p^2)\left(f_+^{(0)}-f_-^{(0)}\right) - \frac\hbar4\epsilon^{\mu\nu\rho\sigma}p_\nu\vp_{\rho\sigma}\,\delta(p^2)\nonumber\\
&\times\!\sum_\chi f_\chi^{(0)}\!\left(1-f_\chi^{(0)}\!\right)\!+\hbar Q\,\Ft^{\mu\nu}p_\nu\,\delta'(p^2)\sum_\chi f_\chi^{(0)},
\label{eq:Aeq}
\end{align}
where $f^{(0)}_\chi\equiv\left[e^{(u\cdot p-\mu_\chi)/T}+1\right]^{-1}$. The vortical term is the massless limit of the global-equilibrium spin polarization of massive fermions~\cite{Becattini:2013fla, Fang:2016vpj} used to describe $\Lambda$ polarization~\cite{STAR:2017ckg, Becattini:2016gvu}. In the rest frame the mean polarization of particles with momentum $\bm p$ is
\begin{equation}
\bm P(\bm p)=\frac{\hbar}{2T}\Big(\bm\omega+\hat{\bm p}\times\bm a+\frac{Q\bm B}{|\bm p|}\Big)\langle1-f\rangle ,
\label{eq:Prest}
\end{equation}
with $\langle1-f\rangle=\sum_\chi f_\chi^{(0)}(1-f_\chi^{(0)})/\sum_\chi f_\chi^{(0)}$. For $\mu_5\neq0$ a helicity term $\hat{\bm p}(f_+^{(0)}-f_-^{(0)})/\sum_\chi f_\chi^{(0)}$ is added. Antiparticles follow from $Q\to-Q$ and $f\to\bar f$. Vorticity and acceleration induce the same polarization for particles and antiparticles, whereas the magnetic field generates an opposite polarization, leading to a particle–antiparticle splitting. In the nondegenerate limit, this difference is approximately $\bm P-\bar{\bm P}\simeq\hbar Q\bm B/(|\bm p|T)$. This splitting is the chiral-fermion counterpart of the $\Lambda$--$\bar\Lambda$ polarization splitting proposed as a magnetometer for heavy-ion collisions~\cite{Becattini:2016gvu}. Moreover, its $1/|\bm p|$ dependence implies that the splitting is strongest at low momenta.

%%%%%%%%%%%%%%%%%%%%%%%%%%%%%%%%%%%%%%%%%%%%%%%%%%

\medskip

\textit{Magnetovortical thermodynamics:}
Since $f_\chi$ represents the Wigner distribution associated with the global-equilibrium density operator~\cite{Becattini:2012tc, Becattini:2013fla}, the pressure in the local rest frame satisfies
\begin{equation}
dP=s\,dT+\sum_\chi n_\chi d\mu_\chi+\bm J\cdot d\bm\omega+\bm M\cdot d\bm B. 
\end{equation}
Here $\bm M$ denotes the magnetization and $\bm J$ is the angular-momentum density. Equation~\eqref{eq:currents}, together with the Maxwell relation $\partial M_j/\partial\omega_i=\partial J_i/\partial B_j$, then determine all terms in $P$ that are bilinear in $\bm\omega$ and $\bm B$, leaving only the magnetic susceptibility $\chi_B$ undetermined. We obtain the final expression (SM, Sec.~S10),
\begin{equation}
P=P_0+\frac{\hbar^2}{4}\xi_5\,\bm\omega^2+\frac{\hbar^2}{2}\xi_{B5}\,\bm\omega\cdot\bm B+\frac12\chi_B\bm B^2, 
\label{eq:P2}
\end{equation}
where $P_0$ is the zeroth-order equilibrium pressure. Using the thermodynamic relation $\bm M=\partial P/\partial \bm B$ and $\bm J=\partial P/\partial\bm\omega$, we obtain (SM, Sec.~S10),
\begin{equation}
\bm M =\frac{\hbar^2}{2}\xi_{B5}\,\bm\omega+\chi_B\bm B, \quad \bm J=\frac{\hbar^2}{2}\big(\xi_5\,\bm\omega+\xi_{B5}\bm B\big).
\end{equation}
Note that the above expression for $\bm J$ coincides with the spin density in comoving frame $\bm J=\hbar\bm j_5/2$ in Eq.~\eqref{eq:currents}.

Our framework leads to three notable predictions:\\
{\bf (i)  Barnett magnetization~\cite{Barnett:1915}:} In covariant form,
\begin{equation}
M^{\mu\nu}=-\frac{\hbar^2T}{2}\,\xi_{B5}\,\Delta^\mu_{\ \alpha}\Delta^\nu_{\ \beta}\,\vp^{\alpha\beta}+\chi_BF^{\mu\nu},
\label{eq:M}
\end{equation}
for $E^\mu=0$. In the comoving frame, the corresponding spatial magnetization is $\bm M=\hbar^2Q\mu\,\bm\omega/(4\pi^2)$, apart from the field contribution proportional to $\chi_{B}$. The ratio of the Barnett magnetization to the rotation-induced spin density is $(|{\bm M}|/|{\bm J}|)_{{\bm B}\to0}=\xi_{B5}/\xi_5$. In the degenerate limit ($T\to0$), this ratio approaches $Q/\mu$ for $\mu_5=0$, corresponding to the gyromagnetic ratio ($g=2$) of a massless fermion at the Fermi surface.\\ 
{\bf (ii) Anomalous charge redistribution:} The interplay of vorticity and magnetic field induces a charge-density correction $\delta n=\hbar^2Q\,\bm\omega\cdot\bm B/(4\pi^2)$. which is independent of temperature and chemical potentials. This result coincides with the lowest-Landau-level prediction of Hattori and Yin~\cite{Hattori:2016njk}, while here it emerges within the weak-field kinetic framework. In noncentral heavy-ion collisions, where $\bm B$ and $\bm\omega$ are approximately aligned, this effect provides a mechanism for electric-charge redistribution.\\
{\bf (iii) Rotation-induced density corrections:} Rotation also modifies the thermodynamic densities according to $\delta n=\hbar^2\mu\bm\omega^2/(4\pi^2)$, $\delta n_5=\hbar^2\mu_5\bm\omega^2/(4\pi^2)$, and $\delta\varepsilon=\frac34\hbar^2\xi_5\bm\omega^2$. For $\mu_5=0$ these corrections reproduce the exact second-order coefficients obtained for the free Dirac field~\cite{Buzzegoli:2017cqy}, providing an independent consistency check of the identification $\bm J=\hbar\bm j_5/2$. At the microscopic level, the magnetically shifted mass shell in Eq.~\eqref{eq:CKE} assigns each fermion a magnetic moment $\bm m_\chi=\chi\hbar Q\hat{\bm p}/(2|\bm p|)$, which is aligned with its spin for  $Q>0$. The dipole moments of the polarized quasiparticles account for one-third of $\bm M$. The remaining two-thirds arise from the Berry-curvature correction to the phase-space density and the anomalous velocity (SM, Sec.~S10).

The magnetic susceptibility $\chi_B$ requires second-order CKT~\cite{Gao:2015gyn, Koshino:2016hiz}. The contribution arising from the Zeeman polarization of magnetic moments is (SM, Sec.~S10),
\begin{equation}
\chi_B^{\rm Z}=\frac{\hbar^2Q^2}{24\pi^2}\sum_\chi\Big[f_\chi(0)+\bar f_\chi(0)\Big]=\frac{\hbar^2Q^2}{12\pi^2}.
\end{equation}
The inverse-momentum dependence of the magnetic moment, $\bm m_\chi\propto1/|\bm p|$, makes this contribution sensitive to the infrared boundary $|\bm p|\to0$, where $f+\bar f=1$ for any $T$ and $\mu$. Consequently, the resulting contribution is independent of temperature and chemical potential, unlike the Curie-law susceptibility characteristic of particles with momentum-independent magnetic moments, such as massive fermions~\cite{Bhadury:2022ulr, Buzzegoli:2022qrr}. Instead, we obtain a Pauli-type paramagnetic contribution independent of temperature and chemical potential. Furthermore, charge-conjugation symmetry requires the magnetic susceptibility to be even in the chemical potential, excluding terms odd in $\mu$.

%%%%%%%%%%%%%%%%%%%%%%%%%%%%%%%%%%%%%%%%%%%%%%%%%%

\medskip

\textit{Summary and outlook:}
We have established a covariant formulation of first-order CKT for chiral fermions in electromagnetic fields, in which the chiral kinetic equation takes the form of a pure streaming equation on a magnetically shifted mass shell, Eqs.~\eqref{eq:divJ}--\eqref{eq:CKE}. The resulting streaming operator obeys the chain rule, elucidating the underlying Liouville structure and providing a natural construction of the kinetic entropy current. Its divergence reduces to the collision term weighted by $\ln[(1-f)/f]$, thereby establishing the kinetic $H$-theorem and ensuring vanishing entropy production in global equilibrium. Building on this structure, we obtain an equilibrium Wigner function independent of the spin frame and a complete set of constitutive relations for relativistic spin magnetohydrodynamics, exact in $\mu/T$. These include the anomalous vector and axial currents, the energy--momentum and spin tensors, and the entropy current, together with the spin polarization of chiral fermions and its magnetic particle--antiparticle splitting. Thermodynamic consistency further determines the second-order magnetovortical responses, including the relativistic Barnett effect, anomalous charge redistribution, and rotation-induced density corrections, in agreement with the exact field-theory results.

Beyond providing a unified microscopic description of these equilibrium phenomena, our formulation establishes the thermodynamic foundation for extending chiral spin magnetohydrodynamics to dissipative, non-equilibrium regimes. In particular, Eq.~\eqref{eq:dS} provides an entropy-production constraint on the admissible collision terms and macroscopic constitutive relations. This opens a systematic route to studying the coupled evolution of spin, vorticity, and electromagnetic fields in relativistic matter, with potential applications ranging from heavy-ion collisions to Weyl semimetals and other quantum materials~\cite{Armitage:2017cjs, RevModPhys.87.1213, Gorini_2024}.

%%%%%%%%%%%%%%%%%%%%%%%%%%%%%%%%%%%%%%%%%%%%%%%%%%

\begin{acknowledgments}
G.M. acknowledges useful discussions with Dhanush. The authors thank Sourav Dey for helpful comments. The authors gratefully acknowledge the Department of Atomic Energy (DAE), India, for financial support.
\end{acknowledgments}

\bibliography{ref}

%%%%%%%%%%%%%%%%%%%%%%%%%%%%%%%%%%%%%%%%%%%%%%%%%%%%%%%%%%%%%%%%%%%%%%%%%%%%%%%
\clearpage
\onecolumngrid
\setcounter{equation}{0}
\setcounter{figure}{0}
\setcounter{table}{0}
\setcounter{section}{0}
\setcounter{secnumdepth}{2}
\setcounter{page}{1}
\renewcommand{\theequation}{S\arabic{equation}}
\renewcommand{\thefigure}{S\arabic{figure}}
\renewcommand{\thesection}{S\arabic{section}}
\renewcommand{\thesubsection}{\thesection.\arabic{subsection}}
\makeatletter
\renewcommand{\p@subsection}{}
\makeatother

\begin{center}
{\large\bfseries Supplemental Material}\\[0.3cm]
{\bfseries Equilibrium relativistic spin magnetohydrodynamics from chiral kinetic theory}\\[0.2cm]
Gopal Mistry and Amaresh Jaiswal\\[0.1cm]
{\itshape School of Physical Sciences, National Institute of Science Education and Research,\\ An OCC of Homi Bhabha National Institute, Jatni-752050, India}
\end{center}

\noindent
The Supplemental Material follows the logic of the Letter. Section~\ref{sec:S1} sets up the chiral Wigner-function equations, and Sec.~\ref{sec:S2} solves them to first order in $\hbar$. Section~\ref{sec:S3} derives the chiral kinetic equation (CKE) in streaming form and its three-dimensional limit. Section~\ref{sec:S4} constructs the entropy current and proves the $H$-theorem. Section~\ref{sec:S5} solves the CKE for global equilibrium. Section~\ref{sec:S6} collects the momentum-integration rules, which are used in Secs.~\ref{sec:S7} and \ref{sec:S8} to compute the currents, the energy-momentum and spin tensors, and the entropy current. Section~\ref{sec:S9} derives the spin polarization. Section~\ref{sec:S10} derives the second-order magnetovortical responses, including the magnetization, from thermodynamic consistency. Equation numbers without the prefix S refer to the Letter. Conventions are those of the Letter: $g^{\mu\nu}={\rm diag}(1,-1,-1,-1)$, $\epsilon^{0123}=-\epsilon_{0123}=+1$, $\Ft^{\mu\nu}=\frac12\epsilon^{\mu\nu\alpha\beta}F_{\alpha\beta}$, $\partial_\mu=\partial/\partial x^\mu$, $\partial_p^\mu=\partial/\partial p_\mu$, and $\nabla_\mu=\partial_\mu-QF_{\mu\nu}\partial_p^\nu$. Momentum integrals are written in units $\hbar=1$, with $\hbar$ kept only as the counting parameter of the semiclassical expansion. In the comoving frame, $\bm B$ and $\bm\omega$ denote the spatial parts of $B^\mu=\Ft^{\mu\nu}u_\nu$ and $\omega^\mu$; with our conventions they coincide with the magnetic field and the angular velocity of the fluid.

%%%%%%%%%%%%%%%%%%%%%%%%%%%%%%%%%%%%%%%%%%%%%%%%%%%%%%%%%%%%%%%%%%%%%%%%%%%%%%%
\section{Wigner-function equations for chiral currents}
\label{sec:S1}

The gauge-covariant Wigner function of a Dirac field $\psi$ with charge $Q$ is
\begin{equation}
W_{\alpha\beta}(x,p)=\int d^4y\; e^{-ip\cdot y/\hbar}\,\big\langle\bar\psi_\beta(x+\tfrac y2)\,U(x+\tfrac y2,x-\tfrac y2)\,\psi_\alpha(x-\tfrac y2)\big\rangle ,
\label{S:W}
\end{equation}
where $U$ is the straight gauge link. For massless fermions in a background field it satisfies Eq.~\eqref{eq:wigner}~\cite{Vasak:1987um,Elze:1986hq,Elze:1986qd}. Higher-order terms in the gradient expansion of the field-dependent operators are of order $\hbar^2\partial^2F$ and do not affect the first-order analysis. With the Clifford decomposition
\begin{equation}
W=\frac14\left[\mathcal F+i\gamma^5\mathcal P+\gamma^\mu V_\mu+\gamma^5\gamma^\mu A_\mu+\frac12\sigma^{\mu\nu}S_{\mu\nu}\right],\qquad \sigma^{\mu\nu}=\frac i2\left[\gamma^\mu,\gamma^\nu\right],
\label{S:Clifford}
\end{equation}
the real and imaginary parts of Eq.~\eqref{eq:wigner} split into two decoupled sets. For $m=0$ the vector and axial-vector components close among themselves~\cite{Vasak:1987um,Zhuang:1995pd,Florkowski:1995ei,Weickgenannt:2019dks,Gao:2020pfu}. In terms of the chiral components $J_\chi^\mu=\frac12(V^\mu+\chi A^\mu)$ they read
\begin{align}
&p_\mu J^\mu_\chi=0, \label{S:c1}\\
&\nabla_\mu J^\mu_\chi=0, \label{S:c2}\\
&p^\mu J^\nu_\chi-p^\nu J^\mu_\chi=-\frac{\chi\hbar}{2}\,\epsilon^{\mu\nu\rho\sigma}\nabla_\rho J_{\chi\sigma}, \label{S:c3}
\end{align}
which are Eqs.~\eqref{eq:Jeqs}. Equation~\eqref{S:c2} is the kinetic equation proper. Equations~\eqref{S:c1} and \eqref{S:c3} are constraints that determine the off-shell structure of $J^\mu_\chi$ order by order in $\hbar$.

%%%%%%%%%%%%%%%%%%%%%%%%%%%%%%%%%%%%%%%%%%%%%%%%%%%%%%%%%%%%%%%%%%%%%%%%%%%%%%%
\section{First-order solution of the constraint equations}
\label{sec:S2}

We expand $J^\mu_\chi=J^{(0)\mu}_\chi+\hbar J^{(1)\mu}_\chi+\mathcal O(\hbar^2)$ and suppress the index $\chi$ on $J$ when no confusion arises.

\paragraph{Zeroth order.}
At $\mathcal O(\hbar^0)$, Eq.~\eqref{S:c3} gives $p^\mu J^{(0)\nu}=p^\nu J^{(0)\mu}$, hence $J^{(0)\mu}=p^\mu\varphi$. Equation~\eqref{S:c1} then requires $p^2\varphi=0$, whose general solution is $\varphi=f^{(0)}_\chi\,\delta(p^2)$:
\begin{equation}
J^{(0)\mu}=p^\mu f^{(0)}_\chi\,\delta(p^2),\qquad
\nabla_\mu J^{(0)\mu}=\delta(p^2)\,p\cdot\nabla f^{(0)}_\chi=0 .
\label{S:J0}
\end{equation}
In Eq.~\eqref{S:J0} we used $F_{\mu\lambda}\partial_p^\lambda(p^\mu\delta(p^2))=F_{\mu\lambda}(g^{\lambda\mu}\delta+2p^\mu p^\lambda\delta')=0$.

\paragraph{First order.}
At $\mathcal O(\hbar)$, Eqs.~\eqref{S:c1} and \eqref{S:c3} become
\begin{equation}
p_\mu J^{(1)\mu}=0,\qquad
p^\mu J^{(1)\nu}-p^\nu J^{(1)\mu}=-\frac{\chi}{2}\,\epsilon^{\mu\nu\rho\sigma}\nabla_\rho J^{(0)}_\sigma .
\label{S:first}
\end{equation}
Using $\nabla_\rho\delta(p^2)=-2QF_{\rho\lambda}p^\lambda\delta'(p^2)$, one finds
\begin{equation}
\nabla_\rho J^{(0)}_\sigma=p_\sigma(\nabla_\rho f^{(0)})\,\delta(p^2)-QF_{\rho\sigma}f^{(0)}\delta(p^2)-2QF_{\rho\lambda}p^\lambda p_\sigma f^{(0)}\delta'(p^2).
\label{S:gradJ0}
\end{equation}
Contracting with $\epsilon^{\mu\nu\rho\sigma}$ requires the Schouten identity
\begin{equation}
\epsilon^{\mu\nu\rho\sigma}F_{\rho\lambda}p^\lambda p_\sigma=p^\mu\Ft^{\nu\lambda}p_\lambda-p^\nu\Ft^{\mu\lambda}p_\lambda+p^2\,\Ft^{\mu\nu},
\label{S:schoutenA}
\end{equation}
which follows from antisymmetrizing over five indices in four dimensions. With $\epsilon^{\mu\nu\rho\sigma}F_{\rho\sigma}=2\Ft^{\mu\nu}$ and the distributional identity $p^2\delta'(p^2)=-\delta(p^2)$, the two terms proportional to $\Ft^{\mu\nu}\delta(p^2)$ cancel and
\begin{equation}
\epsilon^{\mu\nu\rho\sigma}\nabla_\rho J^{(0)}_\sigma=\epsilon^{\mu\nu\rho\sigma}p_\sigma(\nabla_\rho f^{(0)})\,\delta(p^2)-2Q\big(p^\mu\Ft^{\nu\lambda}-p^\nu\Ft^{\mu\lambda}\big)p_\lambda f^{(0)}\delta'(p^2).
\label{S:epsgradJ0}
\end{equation}
The second term follows from the contribution $\chi Q\Ft^{\mu\lambda}p_\lambda f^{(0)}\delta'(p^2)$ to $J^{(1)\mu}$. For the first term we introduce a timelike frame vector $n^\mu$ and the spin tensor $S^{\mu\nu}_{(n)}=\epsilon^{\mu\nu\rho\sigma}p_\rho n_\sigma/(2\,n\cdot p)$. For $p^2=0$ and any $v_\rho$ with $p\cdot v=0$, the Schouten identity gives
\begin{equation}
p^\mu S^{\nu\rho}_{(n)}v_\rho-p^\nu S^{\mu\rho}_{(n)}v_\rho=-\frac12\,\epsilon^{\mu\nu\rho\sigma}v_\rho p_\sigma .
\label{S:schoutenB}
\end{equation}
Applied to $v_\rho=\nabla_\rho f^{(0)}$, which satisfies $p\cdot v=0$ on shell by Eq.~\eqref{S:J0}. This confirms that the side-jump contribution $\chi S^{\mu\nu}_{(n)}(\nabla_\nu f^{(0)})\delta(p^2)$ to $J^{(1)\mu}$ reproduces the first term of Eq.~\eqref{S:epsgradJ0}. Since $p_\mu S^{\mu\nu}_{(n)}=0$ and $p_\mu\Ft^{\mu\nu}p_\nu=0$, both pieces obey $p_\mu J^{(1)\mu}=0$. The homogeneous solution of Eq.~\eqref{S:first} is $p^\mu f^{(1)}_\chi\delta(p^2)$. Collecting terms and defining $f_\chi=f^{(0)}_\chi+\hbar f^{(1)}_\chi$, we obtain Eq.~\eqref{eq:J},
\begin{equation}
J^\mu_\chi=\delta(p^2)\big[p^\mu+\chi\hbar\,S^{\mu\nu}_{(n)}\nabla_\nu\big]f_\chi+\chi\hbar Q\,\Ft^{\mu\nu}p_\nu\,\delta'(p^2)\,f_\chi+\mathcal O(\hbar^2).
\label{S:J}
\end{equation}

\paragraph{Frame dependence.}
For two frame vectors $n$ and $n'$, the Schouten identity gives, for $p^2=0$ and $p\cdot v=0$,
\begin{equation}
\big(S^{\mu\nu}_{(n)}-S^{\mu\nu}_{(n')}\big)v_\nu=p^\mu\,\frac{\epsilon^{\alpha\beta\gamma\delta}p_\alpha n_\beta n'_\gamma v_\delta}{2(n\cdot p)(n'\cdot p)} .
\label{S:frame}
\end{equation}
The $\delta'(p^2)$ term in Eq.~\eqref{S:J} does not involve $n$. Hence $J^\mu_\chi$ is frame independent if the distribution functions are related by
\begin{equation}
f_{\chi(n')}=f_{\chi(n)}+\chi\hbar\,\frac{\epsilon^{\alpha\beta\gamma\delta}p_\alpha n_\beta n'_\gamma\nabla_\delta f_\chi}{2(n\cdot p)(n'\cdot p)}+\mathcal O(\hbar^2).
\label{S:ftrans}
\end{equation}
The arbitrariness of $n^\mu$ reflects the spin supplementary condition for massless particles~\cite{Chen:2014cla,Hidaka:2016yjf,Costa:2017kdr}.

%%%%%%%%%%%%%%%%%%%%%%%%%%%%%%%%%%%%%%%%%%%%%%%%%%%%%%%%%%%%%%%%%%%%%%%%%%%%%%%
\section{Chiral kinetic equation in streaming form}
\label{sec:S3}

\subsection{Divergence of the current functional}
Let $J^\mu_\chi[h]$ denote the right-hand side of Eq.~\eqref{S:J} with $f_\chi$ replaced by an arbitrary function $h(x,p)$. We compute $\nabla_\mu J^\mu_\chi[h]$ term by term without using any equation of motion.

(i) As in Eq.~\eqref{S:J0}, $\nabla_\mu[p^\mu h\,\delta(p^2)]=\delta(p^2)\,p\cdot\nabla h$.

(ii) For the side-jump term,
\begin{equation}
\nabla_\mu\big[S^{\mu\nu}_{(n)}(\nabla_\nu h)\delta(p^2)\big]=\delta(p^2)\,\nabla_\mu\big(S^{\mu\nu}_{(n)}\nabla_\nu h\big)-2QF_{\mu\lambda}p^\lambda S^{\mu\nu}_{(n)}(\nabla_\nu h)\,\delta'(p^2).
\label{S:ii}
\end{equation}

(iii) For the magnetic-moment term, the Bianchi identity $\partial_\mu\Ft^{\mu\nu}=0$, the relation $F_{\mu\lambda}\Ft^{\mu\nu}=\frac14\delta^\nu_\lambda F_{\alpha\beta}\Ft^{\alpha\beta}$, and $p^2\delta''(p^2)=-2\delta'(p^2)$ show that the terms in which $\nabla_\mu$ acts on $\Ft^{\mu\nu}p_\nu\delta'(p^2)$ cancel:
\begin{equation}
\nabla_\mu\big[\Ft^{\mu\nu}p_\nu\,\delta'(p^2)\,h\big]=\Ft^{\mu\nu}p_\nu\,(\nabla_\mu h)\,\delta'(p^2).
\label{S:iii}
\end{equation}

(iv) The Schouten identity~\eqref{S:schoutenA}, contracted with $n_\nu$, gives the algebraic identity
$(n\cdot p)\Ft^{\mu\nu}p_\nu=p^\mu\,\Ft^{\alpha\beta}n_\alpha p_\beta+p^2B^\mu_{(n)}-2(n\cdot p)S^{\mu\alpha}_{(n)}F_{\alpha\lambda}p^\lambda$, with $B^\mu_{(n)}=\Ft^{\mu\nu}n_\nu$. Using $\Ft^{\alpha\beta}n_\alpha p_\beta=-(n\cdot p)F_{\alpha\beta}S^{\alpha\beta}_{(n)}$ and $p^2\delta'=-\delta$, it becomes
\begin{equation}
\Ft^{\mu\nu}p_\nu\,\delta'(p^2)=-\big(F_{\alpha\beta}S^{\alpha\beta}_{(n)}\big)\,p^\mu\,\delta'(p^2)-\frac{B^\mu_{(n)}}{n\cdot p}\,\delta(p^2)-2S^{\mu\alpha}_{(n)}F_{\alpha\lambda}p^\lambda\,\delta'(p^2).
\label{S:iv}
\end{equation}

(v) Finally, $\nabla_\mu(S^{\mu\nu}_{(n)}\nabla_\nu h)=(\nabla_\mu S^{\mu\nu}_{(n)})\nabla_\nu h+\frac12S^{\mu\nu}_{(n)}[\nabla_\mu,\nabla_\nu]h$, with
\begin{equation}
-QF_{\mu\lambda}\partial_p^\lambda S^{\mu\nu}_{(n)}=\frac{Q}{n\cdot p}\big(B^\nu_{(n)}+E_\mu S^{\mu\nu}_{(n)}\big),\qquad
[\nabla_\mu,\nabla_\nu]=-Q(\partial_\mu F_{\nu\lambda}-\partial_\nu F_{\mu\lambda})\partial_p^\lambda,
\label{S:v}
\end{equation}
where $E_\mu=F_{\mu\lambda}n^\lambda$.

We now assemble the pieces. The term $\propto S^{\mu\alpha}_{(n)}F_{\alpha\lambda}p^\lambda\delta'(p^2)$ generated in (iv) cancels the last term of Eq.~\eqref{S:ii}. The term $-B^\nu_{(n)}\nabla_\nu h/(n\cdot p)$ from (iv) cancels the $B^\nu_{(n)}$ term in Eq.~\eqref{S:v}. The remaining $\delta'$ term, $-\chi\hbar Q(F_{\alpha\beta}S^{\alpha\beta}_{(n)})\,\delta'(p^2)\,p\cdot\nabla h$, combines with $\delta(p^2)\,p\cdot\nabla h$ into a shifted delta function. The result is
\begin{equation}
\nabla_\mu J^\mu_\chi[h]=\delta\big(p^2-\chi\hbar QF_{\alpha\beta}S^{\alpha\beta}_{(n)}\big)\,\Dc h+\mathcal O(\hbar^2),
\label{S:divJ}
\end{equation}
\begin{equation}
\Dc=p^\mu\nabla_\mu+\chi\hbar\Big[\partial_\mu S^{\mu\nu}_{(n)}+\frac{Q}{n\cdot p}E_\mu S^{\mu\nu}_{(n)}\Big]\nabla_\nu-\chi\hbar Q\,S^{\mu\nu}_{(n)}(\partial_\mu F_{\nu\lambda})\,\partial_p^\lambda .
\label{S:Dop}
\end{equation}
These are Eqs.~\eqref{eq:divJ} and \eqref{eq:D}. Since $F_{\alpha\beta}S^{\alpha\beta}_{(n)}=B_{(n)}\cdot p/(n\cdot p)$, the shell is $p^2=\chi\hbar Q\,B_{(n)}\cdot p/(n\cdot p)$. Distributions $\sum_kc_k(x,p)\,\delta^{(k)}(p^2)$ were reduced to a canonical form on the shell, for a generic polynomial $h(x,p)$, a position-dependent frame $n_\mu(x)$, and a position-dependent field $F_{\mu\nu}=\partial_\mu A_\nu-\partial_\nu A_\mu$; the two sides agree identically at $\mathcal O(\hbar)$ for both chiralities.

\paragraph{Properties.}
(a) $\Dc$ is a first-order differential operator in phase space without a multiplicative term. Hence $\Dc\Phi(h)=\Phi'(h)\,\Dc h$ and $\Dc 1=0$. This Leibniz property underlies the $H$-theorem of Sec.~\ref{sec:S4}. It is lost if the $\delta'(p^2)$ terms are converted into momentum derivatives of $f_\chi$ by integrating by parts: a multiplicative term then appears, which mimics a compressible phase-space flow.
(b) For constant $F_{\mu\nu}$ and constant $n^\mu$, $\Dc=p\cdot\nabla+\chi\hbar QE_\mu S^{\mu\nu}_{(n)}\nabla_\nu/(n\cdot p)$. If in addition $E_\mu=0$, then $\Dc=p\cdot\nabla$ and all Berry-curvature effects reside in the shell.
(c) Integrating Eq.~\eqref{S:divJ} and its antiparticle counterpart over the positive-energy branch, Eq.~\eqref{S:moments}, gives $\partial_\mu j^\mu_\chi=QF_{\mu\lambda}\int_P\partial_p^\lambda\big(J^\mu_\chi+\bar J^\mu_\chi\big)$. This total derivative receives a contribution only from the tip $p^\mu=0$ of the cone, where the Berry monopole sits, and reproduces the chiral anomaly $\propto\chi Q^2F_{\alpha\beta}\Ft^{\alpha\beta}$~\cite{Son:2012wh,Hidaka:2016yjf}. Particles and antiparticles contribute in proportion to $f_\chi(0)$ and $\bar f_\chi(0)$, whose sum is unity in equilibrium, so the coefficient is that of field theory. The anomaly vanishes for the configurations considered here, $F_{\alpha\beta}\Ft^{\alpha\beta}=0$.

\subsection{Three-dimensional form and Liouville theorem}
Take $n^\mu=(1,\bm 0)$ and a constant field with $\bm E=0$. Then $B_{(n)}\cdot p=-\bm B\cdot\bm p$, and the shell $p^2=\chi\hbar Q\,B_{(n)}\cdot p/p^0$ gives
\begin{equation}
p^0=E_{\bm p}=|\bm p|-\chi\hbar Q\,\frac{\bm B\cdot\hat{\bm p}}{2|\bm p|}=|\bm p|\,\big(1-Q\bm B\cdot\bm\Omega_\chi\big),\qquad
\bm\Omega_\chi=\chi\hbar\,\frac{\hat{\bm p}}{2|\bm p|^2}.
\label{S:Ep}
\end{equation}
Equation~\eqref{S:Ep} is the magnetic-moment energy of a Weyl fermion. With $F^{ij}=-\epsilon^{ijk}B^k$ one finds $p\cdot\nabla=p^0\partial_t+\bm p\cdot\bm\nabla_{\bm x}+Q(\bm p\times\bm B)\cdot\bm\nabla_{\bm p}$ acting on $f(x,p^0,\bm p)$. For the on-shell distribution $f_3(t,\bm x,\bm p)=f(x,E_{\bm p},\bm p)$, $\bm\nabla_{\bm p}f_3=\bm\nabla_{\bm p}f+\tilde{\bm v}\,\partial_{p^0}f$ with
\begin{equation}
\tilde{\bm v}=\frac{\partial E_{\bm p}}{\partial\bm p}=\hat{\bm p}+\chi\hbar Q\,\frac{(\bm B\cdot\hat{\bm p})\hat{\bm p}}{|\bm p|^2}-\chi\hbar Q\frac{\bm B}{2|\bm p|^2}.
\label{S:vtilde}
\end{equation}
Since $(\bm p\times\bm B)\cdot\tilde{\bm v}=0$, the operator $\Dc=p\cdot\nabla$ is tangent to the shell, and
\begin{equation}
\Dc f=E_{\bm p}\Big(\partial_t+\frac{\bm p}{E_{\bm p}}\cdot\bm\nabla_{\bm x}+Q\frac{\bm p\times\bm B}{E_{\bm p}}\cdot\bm\nabla_{\bm p}\Big)f_3 .
\label{S:D3}
\end{equation}
The equations of motion of CKT~\cite{Stephanov:2012ki,Son:2012wh,Son:2012zy},
$\sqrt\gamma\,\dot{\bm x}=\tilde{\bm v}+Q(\tilde{\bm v}\cdot\bm\Omega_\chi)\bm B$ and $\sqrt\gamma\,\dot{\bm p}=Q\tilde{\bm v}\times\bm B$ with $\sqrt\gamma=1+Q\bm B\cdot\bm\Omega_\chi$, give to first order
\begin{equation}
\dot{\bm x}=(1+Q\bm B\cdot\bm\Omega_\chi)\,\hat{\bm p}=\frac{\bm p}{E_{\bm p}},\qquad
\dot{\bm p}=(1+Q\bm B\cdot\bm\Omega_\chi)\,Q\hat{\bm p}\times\bm B=Q\frac{\bm p\times\bm B}{E_{\bm p}} .
\label{S:eom}
\end{equation}
Equation~\eqref{S:D3} is therefore $\Dc f=E_{\bm p}(\partial_t+\dot{\bm x}\cdot\bm\nabla_{\bm x}+\dot{\bm p}\cdot\bm\nabla_{\bm p})f_3$, which is Eq.~\eqref{eq:CKE3}. The phase-space measure follows from the Jacobian of the shifted delta function:
\begin{equation}
\int dp^0\,\theta(p^0)\,\delta\big(p^2-\chi\hbar QF_{\alpha\beta}S^{\alpha\beta}_{(n)}\big)\,G(p)=\frac{\sqrt\gamma}{2E_{\bm p}}\,G(E_{\bm p},\bm p)+\mathcal O(\hbar^2),
\label{S:measure}
\end{equation}
because $\partial_{p^0}(p^2-\chi\hbar QB_{(n)}\cdot p/p^0)=2E_{\bm p}/\sqrt\gamma$ on shell. The phase-space density is thus $\sqrt\gamma f_3$. Liouville's theorem holds identically: $\partial_t\sqrt\gamma=0$, $\bm\nabla_{\bm x}\cdot(\sqrt\gamma\dot{\bm x})=0$, and $\bm\nabla_{\bm p}\cdot(\sqrt\gamma\dot{\bm p})=Q\bm B\cdot(\bm\nabla_{\bm p}\times\tilde{\bm v})=0$. Consequently
\begin{equation}
\partial_t(\sqrt\gamma f_3)+\bm\nabla_{\bm x}\cdot(\sqrt\gamma\dot{\bm x}f_3)+\bm\nabla_{\bm p}\cdot(\sqrt\gamma\dot{\bm p}f_3)=\sqrt\gamma\big(\partial_t+\dot{\bm x}\cdot\bm\nabla_{\bm x}+\dot{\bm p}\cdot\bm\nabla_{\bm p}\big)f_3,
\label{S:Liouville}
\end{equation}
so the conservation law~\eqref{S:c2} and the streaming equation are equivalent, with no compressibility term. For $\bm E\neq0$ the same structure holds away from $\bm p=0$; the monopole $\bm\nabla_{\bm p}\cdot\bm\Omega_\chi=2\pi\chi\hbar\,\delta^3(\bm p)$ gives the anomalous source $\propto Q^2\bm E\cdot\bm B$~\cite{Stephanov:2012ki,Son:2012zy}.

%%%%%%%%%%%%%%%%%%%%%%%%%%%%%%%%%%%%%%%%%%%%%%%%%%%%%%%%%%%%%%%%%%%%%%%%%%%%%%%
\section{Entropy current and \texorpdfstring{$H$}{H}-theorem}
\label{sec:S4}

\paragraph{Antiparticles.}
The constraint equations hold for both signs of $p^0$. For $p^0<0$ the vacuum-subtracted distribution is $f_\chi(x,p)=\bar f_\chi(x,-p)-1$, where $\bar f_\chi$ is the antiparticle distribution. Substituting $p=-\bar p$ in Eq.~\eqref{S:J}, using $S^{\mu\nu}_{(n)}(-\bar p)=S^{\mu\nu}_{(n)}(\bar p)$, $\partial_p=-\partial_{\bar p}$, and dropping the vacuum term, one finds
\begin{equation}
J^\mu_\chi\big|_{p^0<0}(x,-\bar p)=-\bar J^\mu_\chi[\bar f_\chi](x,\bar p),\qquad
\bar J^\mu_\chi[h]\equiv J^\mu_\chi[h]\big|_{(Q,\chi)\to(-Q,-\chi)} .
\label{S:anti}
\end{equation}
Antiparticles of chirality $\chi$ thus move as particles of charge $-Q$ and opposite helicity, as expected.

\paragraph{Entropy current.}
Let $\int_P\equiv2\int d^4p\,\theta(p^0)/(2\pi)^3$ and $\Phi(f)=f\ln f+(1-f)\ln(1-f)$. We define the entropy current as in Eq.~\eqref{eq:Smu}:
\begin{equation}
\mathcal H^\mu=-\sum_\chi\int_P\Big(J^\mu_\chi[\Phi(f_\chi)]+\bar J^\mu_\chi[\Phi(\bar f_\chi)]\Big).
\label{S:Smu}
\end{equation}
At zeroth order this reduces to the familiar $-\int d^3p\,(p^\mu/|\bm p|)\Phi(f)/(2\pi)^3$. At first order it contains the side-jump and magnetic-moment contributions dictated by Eq.~\eqref{S:J}.

\paragraph{Divergence.}
Writing $\partial_\mu=\nabla_\mu+QF_{\mu\lambda}\partial_p^\lambda$, the second term produces a total momentum derivative. For $F_{\alpha\beta}\Ft^{\alpha\beta}=0$ it integrates to zero, since the only possible contribution, from the Berry monopole at $p=0$, is proportional to the anomaly [property (c) of Sec.~\ref{sec:S3}]. Equation~\eqref{S:divJ} applied to $h=\Phi(f_\chi)$, together with the Leibniz rule $\Dc\Phi(f)=\Phi'(f)\Dc f$ and $\Phi'(f)=\ln[f/(1-f)]$, gives
\begin{equation}
\partial_\mu\mathcal H^\mu=\sum_\chi\int_P\delta\big(p^2-\chi\hbar QF_{\alpha\beta}S^{\alpha\beta}_{(n)}\big)\,
\ln\frac{1-f_\chi}{f_\chi}\,\Dc f_\chi+(\text{antiparticles})+\mathcal O(\hbar^2),
\label{S:dS}
\end{equation}
which is Eq.~\eqref{eq:dS}.

\paragraph{Collisionless dynamics and $H$-theorem.}
If $f_\chi$ solves the CKE~\eqref{eq:CKE}, the integrand vanishes pointwise in phase space and $\partial_\mu\mathcal H^\mu=0$. This is the relativistic covariant counterpart of the conservation of the Gibbs entropy under Liouville flow, Eq.~\eqref{S:Liouville}. In the presence of collisions, $\Dc f_\chi=C_\chi[f]$ on the shell, Eq.~\eqref{S:dS} becomes
\begin{equation}
\partial_\mu\mathcal H^\mu=\sum_\chi\int_P\delta\big(p^2-\chi\hbar QF_{\alpha\beta}S^{\alpha\beta}_{(n)}\big)\,\psi_\chi\,C_\chi[f]+(\text{antiparticles}),\qquad
\psi_\chi\equiv\ln\frac{1-f_\chi}{f_\chi}.
\label{S:Htheorem}
\end{equation}
For a collision kernel that conserves charge, energy-momentum, and total angular momentum, including the side-jump nonlocality required by Lorentz covariance~\cite{Chen:2015gta}, the standard symmetrization argument gives $\partial_\mu\mathcal H^\mu\geq0$~\cite{Jaiswal:2013fc}. Equality holds if and only if $\psi_\chi$ is a linear combination of the summational invariants $1$, $p^\mu$, and $J^{\mu\nu}=x^\mu p^\nu-x^\nu p^\mu+\chi\hbar S^{\mu\nu}_{(n)}$. Section~\ref{sec:S5} shows that the equilibrium solution has exactly this form, $\psi_\chi=g_\chi$. Its entropy production therefore vanishes both because $\Dc f^{\rm eq}_\chi=0$ and because $\int\psi_\chi C_\chi=0$.

%%%%%%%%%%%%%%%%%%%%%%%%%%%%%%%%%%%%%%%%%%%%%%%%%%%%%%%%%%%%%%%%%%%%%%%%%%%%%%%
\section{Global equilibrium}
\label{sec:S5}

We look for $f_\chi=[e^{g_\chi}+1]^{-1}$. Since $\Dc f_\chi=-f_\chi(1-f_\chi)\,\Dc g_\chi$, the CKE requires $\Dc g_\chi=0$ on the shell. We write $g_\chi=\beta_\mu p^\mu-\alpha_\chi+\chi\hbar\, g^{(1)}_\chi$.

\paragraph{Zeroth order.}
From $\nabla_\mu(\beta\cdot p)=p^\lambda\partial_\mu\beta_\lambda-QF_{\mu\lambda}\beta^\lambda$,
\begin{equation}
p^\mu\nabla_\mu(\beta\cdot p-\alpha_\chi)=p^\mu p^\nu\partial_{(\mu}\beta_{\nu)}-p^\mu\big(\partial_\mu\alpha_\chi+QF_{\mu\lambda}\beta^\lambda\big).
\label{S:zeroth}
\end{equation}
On the null cone the first term allows a conformal Killing vector, $\partial_{(\mu}\beta_{\nu)}\propto g_{\mu\nu}$. At first order, however, the shell is $p^2=\mathcal O(\hbar)$ and the spin term below requires a constant thermal vorticity. Both conditions select a Killing vector~\cite{Hidaka:2016yjf,Becattini:2012tc,Becattini:2013fla}:
\begin{equation}
\partial_\mu\beta_\nu+\partial_\nu\beta_\mu=0\;\;\Longrightarrow\;\;\beta_\mu=b_\mu+\vp_{\mu\nu}x^\nu,\quad \partial_\mu\beta_\nu=-\vp_{\mu\nu}={\rm const}.
\label{S:Killing}
\end{equation}
The second term in Eq.~\eqref{S:zeroth} vanishes if $\partial_\mu\alpha_\chi=-QF_{\mu\lambda}\beta^\lambda$. Using the Bianchi identity, its integrability condition is
\begin{equation}
\partial_\nu(F_{\mu\lambda}\beta^\lambda)-\partial_\mu(F_{\nu\lambda}\beta^\lambda)=\beta^\lambda\partial_\lambda F_{\mu\nu}+F_{\lambda\nu}\partial_\mu\beta^\lambda+F_{\mu\lambda}\partial_\nu\beta^\lambda\equiv\pounds_\beta F_{\mu\nu}=0,
\label{S:Lie}
\end{equation}
i.e., the field must be invariant along the flow generated by $\beta$. These are Eqs.~\eqref{eq:Killing}. In the comoving frame the electric field $E_\mu=F_{\mu\nu}u^\nu$ is then balanced by the gradient of the chemical potential, $T\partial_\mu(\mu_\chi/T)=-QE_\mu$, i.e., $\bm\nabla(\mu_\chi/T)=Q\bm E/T$. For a rigidly rotating fluid in a constant magnetic field, Eq.~\eqref{S:Lie} requires $\bm B\parallel\bm\omega$, and $\alpha_\chi=\alpha_0+Q B\Omega(x^2+y^2)/(2T_0)$ for rotation with angular velocity $\Omega$ about $\hat{\bm z}$.

\paragraph{First order: frame covariance.}
The spin term is fixed by frame covariance. Under $n\to n'$, $f_\chi$ must change according to Eq.~\eqref{S:ftrans}, with $\nabla_\delta f_\chi=-f_\chi(1-f_\chi)\nabla_\delta g_\chi$ and $\nabla_\delta(\beta\cdot p-\alpha_\chi)=-\vp_{\delta\lambda}p^\lambda$. Equation~\eqref{S:frame} with $v_\nu=-\vp_{\nu\lambda}p^\lambda$ (note $p\cdot v=0$), combined with the identity
\begin{equation}
\big(S^{\mu\nu}_{(n)}-S^{\mu\nu}_{(n')}\big)\vp_{\nu\lambda}p^\lambda=-\frac12\,p^\mu\,\vp_{\alpha\beta}\big(S^{\alpha\beta}_{(n)}-S^{\alpha\beta}_{(n')}\big)\qquad(p^2=0),
\label{S:eqframe}
\end{equation}
shows that frame covariance requires $g^{(1)}_\chi(n')-g^{(1)}_\chi(n)=-\frac12\vp_{\alpha\beta}[S^{\alpha\beta}_{(n')}-S^{\alpha\beta}_{(n)}]$. Hence
\begin{equation}
g_\chi=\beta\cdot p-\alpha_\chi-\frac{\chi\hbar}{2}\,\vp_{\alpha\beta}S^{\alpha\beta}_{(n)} ,
\label{S:geq}
\end{equation}
up to an $n$-independent first-order term. By the $H$-theorem of Sec.~\ref{sec:S4}, such a term must itself be a summational invariant, i.e., a combination of $1$, $p^\mu$, and $J^{\mu\nu}$, and can be absorbed into $\alpha_\chi$ and $\beta_\mu$. In particular, no explicit electromagnetic term enters $g_\chi$: the magnetic-moment energy is part of $p^0$ on the shell, Eq.~\eqref{S:Ep}, and is therefore already contained in $\beta\cdot p$. Writing $\beta_\mu=b_\mu+\vp_{\mu\nu}x^\nu$ gives $g_\chi=b\cdot p-\alpha_\chi-\frac12\vp_{\mu\nu}J^{\mu\nu}$ with $J^{\mu\nu}=x^\mu p^\nu-x^\nu p^\mu+\chi\hbar S^{\mu\nu}_{(n)}$, a combination of summational invariants.

\paragraph{First order: kinetic equation.}
The zeroth-order part of $\Dc g_\chi$, Eq.~\eqref{S:zeroth}, vanishes for all $p$ once Eqs.~\eqref{S:Killing} and \eqref{S:Lie} hold. The $\mathcal O(\hbar)$ shift of the mass shell therefore generates no additional terms, and the $\mathcal O(\hbar)$ part of $\Dc g_\chi=0$ needs to be imposed only at $p^2=0$. With $\nabla_\nu(\beta\cdot p-\alpha_\chi)=p^\lambda\partial_\nu\beta_\lambda-\partial_\nu\alpha_\chi-QF_{\nu\lambda}\beta^\lambda=-\vp_{\nu\lambda}p^\lambda$ and $\partial_p^\lambda(\beta\cdot p-\alpha_\chi)=\beta^\lambda$, it reads
\begin{equation}
p^\mu\nabla_\mu g^{(1)}_\chi-\Big[\partial_\mu S^{\mu\nu}_{(n)}+\frac{Q}{n\cdot p}E_\mu S^{\mu\nu}_{(n)}\Big]\vp_{\nu\lambda}p^\lambda-QS^{\mu\nu}_{(n)}(\partial_\mu F_{\nu\lambda})\beta^\lambda=0,\qquad p^2=0.
\label{S:first_eq}
\end{equation}
We now show that $g^{(1)}_\chi=-\frac12\vp_{\alpha\beta}S^{\alpha\beta}_{(n)}$ satisfies Eq.~\eqref{S:first_eq} identically.

\textit{Notation and tools.} The spin tensor depends on the frame vector only through $m_\sigma\equiv n_\sigma/(n\cdot p)$, which obeys $p\cdot m=1$: $S^{\mu\nu}_{(n)}=\frac12\epsilon^{\mu\nu\rho\sigma}p_\rho m_\sigma$. For an antisymmetric tensor $A_{\mu\nu}$ we write $\widetilde A^{\mu\nu}=\frac12\epsilon^{\mu\nu\alpha\beta}A_{\alpha\beta}$. Then $A_{\alpha\beta}S^{\alpha\beta}_{(n)}=\widetilde A^{\rho\sigma}p_\rho m_\sigma$, and the inverse relation is $A_{\nu\lambda}=-\frac12\epsilon_{\nu\lambda\alpha\beta}\widetilde A^{\alpha\beta}$. We use the contraction $\epsilon^{\mu\nu\rho\sigma}\epsilon_{\nu\lambda\alpha\beta}=\delta^{\mu\rho\sigma}_{\lambda\alpha\beta}$, where $\delta^{\mu\rho\sigma}_{\lambda\alpha\beta}$ is the generalized Kronecker delta (the determinant of the $3\times3$ matrix of Kronecker deltas, with rows $\mu,\rho,\sigma$ and columns $\lambda,\alpha,\beta$). We also use the Schouten identity for a traceless matrix, here $F^\mu_{\ \kappa}=g^{\mu\alpha}F_{\alpha\kappa}$:
\begin{equation}
F^a_{\ \kappa}\epsilon^{\kappa bcd}+F^b_{\ \kappa}\epsilon^{a\kappa cd}+F^c_{\ \kappa}\epsilon^{ab\kappa d}+F^d_{\ \kappa}\epsilon^{abc\kappa}=0 .
\label{S:Sch4}
\end{equation}
For any antisymmetric $A$, the inverse relation and the $\epsilon$ contraction give $S^{\mu\nu}_{(n)}A_{\nu\lambda}p^\lambda=-\frac14\delta^{\mu\rho\sigma}_{\lambda\alpha\beta}\,p^\lambda\widetilde A^{\alpha\beta}p_\rho m_\sigma$. Of the six terms of the generalized delta, two are proportional to $p^2$; with $p\cdot m=1$, the remaining four give
\begin{equation}
S^{\mu\nu}_{(n)}A_{\nu\lambda}p^\lambda=-\frac12\,p^\mu A_{\alpha\beta}S^{\alpha\beta}_{(n)}-\frac12\,\widetilde A^{\mu\nu}p_\nu\qquad(p^2=0).
\label{S:idI}
\end{equation}
For $A=\vp$ this is Eq.~\eqref{S:X}; because the last term does not depend on $n$, it also implies Eq.~\eqref{S:eqframe}.

\textit{Decomposition.} We insert $g^{(1)}_\chi=-\frac12\vp_{\alpha\beta}S^{\alpha\beta}_{(n)}$, use $\partial_\mu\vp_{\alpha\beta}=0$, and write $p^\mu\nabla_\mu=p^\mu\partial_\mu-Qp^\mu F_{\mu\lambda}\partial_p^\lambda$. The left-hand side of Eq.~\eqref{S:first_eq} then becomes $\sum_{i=1}^5\mathcal T_i$ with
\begin{align}
\mathcal T_1&=-\tfrac12\vp_{\alpha\beta}\,p^\mu\partial_\mu S^{\alpha\beta}_{(n)}, &
\mathcal T_2&=\tfrac Q2\,p^\mu F_{\mu\lambda}\,\partial_p^\lambda\big(\vp_{\alpha\beta}S^{\alpha\beta}_{(n)}\big), &
\mathcal T_3&=-\big(\partial_\mu S^{\mu\nu}_{(n)}\big)\vp_{\nu\lambda}p^\lambda, \nonumber\\
\mathcal T_4&=-\frac{Q}{n\cdot p}\,E_\mu S^{\mu\nu}_{(n)}\vp_{\nu\lambda}p^\lambda, &
\mathcal T_5&=-QS^{\mu\nu}_{(n)}(\partial_\mu F_{\nu\lambda})\beta^\lambda .
\label{S:Tterms}
\end{align}
The terms $\mathcal T_1$ and $\mathcal T_3$ arise only for a position-dependent frame vector, and $\mathcal T_5$ only for an inhomogeneous field.

\textit{Step 1: frame gradients.} Let $M_{\mu\sigma}\equiv\partial_\mu m_\sigma$. Since $p$ is independent of $x$, $p^\sigma M_{\mu\sigma}=\partial_\mu(p\cdot m)=0$. Then $\partial_\mu S^{\alpha\beta}_{(n)}=\frac12\epsilon^{\alpha\beta\rho\sigma}p_\rho M_{\mu\sigma}$, so $\mathcal T_1=-\frac12\widetilde\vp^{\rho\sigma}p_\rho\,p^\mu M_{\mu\sigma}$. In $\mathcal T_3$ we write $\vp_{\nu\lambda}p^\lambda=-\frac12\epsilon_{\nu\lambda\alpha\beta}p^\lambda\widetilde\vp^{\alpha\beta}$ and obtain
\begin{equation}
\mathcal T_3=\frac14\,\delta^{\mu\rho\sigma}_{\lambda\alpha\beta}\,M_{\mu\sigma}\,p_\rho\,p^\lambda\,\widetilde\vp^{\alpha\beta}=\frac12\,\widetilde\vp^{\rho\sigma}p_\rho\,p^\mu M_{\mu\sigma}=-\mathcal T_1 .
\label{S:step1}
\end{equation}
Of the six terms of the generalized delta, two are proportional to $p^2$, two contain $p^\sigma M_{\mu\sigma}=0$, and the remaining two are equal. The frame gradients therefore cancel identically, for any $n^\mu(x)$.

\textit{Step 2: field gradients.} By the antisymmetry of $S^{\mu\nu}_{(n)}$ and the Bianchi identity $\partial_\mu F_{\nu\lambda}+\partial_\nu F_{\lambda\mu}+\partial_\lambda F_{\mu\nu}=0$, $S^{\mu\nu}_{(n)}\partial_\mu F_{\nu\lambda}=-\frac12S^{\mu\nu}_{(n)}\partial_\lambda F_{\mu\nu}$, so $\mathcal T_5=\frac Q2S^{\mu\nu}_{(n)}\beta^\lambda\partial_\lambda F_{\mu\nu}$. With $\partial_\mu\beta^\lambda=-\vp_\mu^{\ \lambda}$, the stationarity condition~\eqref{S:Lie} reads $\beta^\lambda\partial_\lambda F_{\mu\nu}=F_{\lambda\nu}\vp_\mu^{\ \lambda}+F_{\mu\lambda}\vp_\nu^{\ \lambda}$. Contracted with $S^{\mu\nu}_{(n)}$, its two terms contribute equally, and
\begin{equation}
\mathcal T_5=Q\,S^{\mu\nu}_{(n)}F_{\mu\lambda}\vp_\nu^{\ \lambda}.
\label{S:step2}
\end{equation}
The field inhomogeneity therefore enters only through the algebraic combination~\eqref{S:step2}.

\textit{Step 3: algebraic terms.} From $\partial_p^\lambda m_\sigma=-m_\sigma n^\lambda/(n\cdot p)$ one finds $\partial_p^\lambda(\vp_{\alpha\beta}S^{\alpha\beta}_{(n)})=\widetilde\vp^{\lambda\sigma}m_\sigma-(\vp_{\alpha\beta}S^{\alpha\beta}_{(n)})\,n^\lambda/(n\cdot p)$. Hence, with $p\cdot E=p^\mu F_{\mu\lambda}n^\lambda$,
\begin{equation*}
\mathcal T_2=\frac{Q}{2(n\cdot p)}\Big[p^\mu F_{\mu\lambda}\widetilde\vp^{\lambda\sigma}n_\sigma-(p\cdot E)\,\vp_{\alpha\beta}S^{\alpha\beta}_{(n)}\Big].
\end{equation*}
Applying Eq.~\eqref{S:idI} with $A=\vp$ to $\mathcal T_4$ gives
\begin{equation*}
\mathcal T_4=\frac{Q}{2(n\cdot p)}\Big[(p\cdot E)\,\vp_{\alpha\beta}S^{\alpha\beta}_{(n)}+E_\mu\widetilde\vp^{\mu\nu}p_\nu\Big].
\end{equation*}
The terms proportional to $p\cdot E$ cancel, leaving
\begin{equation}
\mathcal T_2+\mathcal T_4=\frac{Q}{2(n\cdot p)}\Big[p^\mu F_{\mu\lambda}\widetilde\vp^{\lambda\sigma}n_\sigma+n^\kappa F_{\mu\kappa}\widetilde\vp^{\mu\nu}p_\nu\Big].
\label{S:T24}
\end{equation}
Now contract the Schouten identity~\eqref{S:Sch4} with $\vp_{cd}\,p_a n_b$. The first two terms give twice the two terms in the bracket of Eq.~\eqref{S:T24}. The last two terms are equal and give $2\epsilon^{ab\mu\nu}p_a n_bF_{\mu\lambda}\vp_\nu^{\ \lambda}$. Therefore, for arbitrary $p$ and $n$,
\begin{equation}
p^\mu F_{\mu\lambda}\widetilde\vp^{\lambda\sigma}n_\sigma+n^\kappa F_{\mu\kappa}\widetilde\vp^{\mu\nu}p_\nu=-\epsilon^{\mu\nu\rho\sigma}p_\rho n_\sigma F_{\mu\lambda}\vp_\nu^{\ \lambda}=-2(n\cdot p)\,S^{\mu\nu}_{(n)}F_{\mu\lambda}\vp_\nu^{\ \lambda},
\label{S:step3}
\end{equation}
and $\mathcal T_2+\mathcal T_4=-\mathcal T_5$.

\textit{Conclusion.} Steps 1--3 give $\sum_{i=1}^5\mathcal T_i=0$. Thus $g^{(1)}_\chi=-\frac12\vp_{\alpha\beta}S^{\alpha\beta}_{(n)}$ solves Eq.~\eqref{S:first_eq} for any frame-vector field $n^\mu(x)$, any constant thermal vorticity, and any field with $\pounds_\beta F=0$. Steps 1 and 3 are identities; the stationarity of the field enters only in Step 2. For a constant field, $\mathcal T_5=0$, and Eq.~\eqref{S:Lie} reduces to $F_{\lambda\nu}\vp_\mu^{\ \lambda}+F_{\mu\lambda}\vp_\nu^{\ \lambda}=0$. Then $F_{\mu\lambda}\vp_\nu^{\ \lambda}$ is symmetric in $\mu\nu$, so the right-hand side of Eq.~\eqref{S:step3} vanishes and $\mathcal T_2+\mathcal T_4=0$ by itself. For $\bm B\perp\bm\omega$ this condition fails, $\mathcal T_2+\mathcal T_4\neq0$, and no global equilibrium exists, consistent with the absence of a solution for $\alpha_\chi$. The antiparticle solution follows from Eq.~\eqref{S:anti}: $\bar g_\chi=\beta\cdot p+\alpha_\chi+\frac{\chi\hbar}{2}\vp_{\alpha\beta}S^{\alpha\beta}_{(n)}$.

\paragraph{Comoving frame.}
We set $\beta^\mu=u^\mu/T$, $\alpha_\chi=\mu_\chi/T$, and $n^\mu=u^\mu$, and decompose
\begin{equation}
\vp^{\mu\nu}=\frac1T\epsilon^{\mu\nu\alpha\beta}\omega_\alpha u_\beta+\frac1T\big(a^\mu u^\nu-a^\nu u^\mu\big),\qquad
\omega^\mu=-\frac T2\epsilon^{\mu\nu\rho\sigma}u_\nu\vp_{\rho\sigma}=\frac12\epsilon^{\mu\nu\rho\sigma}u_\nu\partial_\rho u_\sigma,\quad a^\mu=T\vp^{\mu\nu}u_\nu .
\label{S:vpdecomp}
\end{equation}
The Killing condition implies $u\cdot\partial T=0$, $\partial\cdot u=0$, $a^\mu=u\cdot\partial u^\mu$, $\partial_\mu T=Ta_\mu$, and $\partial_\mu u_\nu=a_\mu u_\nu-T\vp_{\mu\nu}$. Using $\vp_{\alpha\beta}S^{\alpha\beta}_{(u)}=-\omega\cdot p/(T\,u\cdot p)$, the equilibrium distribution and its expansion are
\begin{equation}
f_\chi=\bigg[\exp\Big(\frac{u\cdot p-\mu_\chi}{T}+\frac{\chi\hbar}{2T}\frac{\omega\cdot p}{u\cdot p}\Big)+1\bigg]^{-1}
=f^{(0)}_\chi-\frac{\chi\hbar}{2T}\frac{\omega\cdot p}{u\cdot p}\,f^{(0)}_\chi\big(1-f^{(0)}_\chi\big)+\mathcal O(\hbar^2),
\label{S:fcomoving}
\end{equation}
with $f^{(0)}_\chi=[e^{(u\cdot p-\mu_\chi)/T}+1]^{-1}$. In the local rest frame, $\omega\cdot p=-\bm\omega\cdot\bm p$, and on the shell $u\cdot p=E_{\bm p}$, Eq.~\eqref{S:Ep}. This yields Eq.~\eqref{eq:frest} of the Letter, with the measure $\sqrt\gamma$ of Eq.~\eqref{S:measure}. If one prefers to write the distribution on the null cone $p^0=|\bm p|$, the exponent becomes
\begin{equation}
\frac{u\cdot p-\mu_\chi}{T}+\frac{\chi\hbar}{2T}\Big[\frac{\omega\cdot p}{u\cdot p}+Q\frac{B\cdot p}{(u\cdot p)^2}\Big],
\label{S:nullexp}
\end{equation}
where the magnetic term enters with a positive sign; it describes the energy shift $-\bm m_\chi\cdot\bm B$ with $\bm m_\chi=\chi\hbar Q\hat{\bm p}/(2|\bm p|)$. Moments computed this way must include the Jacobian $\sqrt\gamma$. Equivalently, one may keep $\delta(p^2)$ and the explicit $\delta'(p^2)$ term of Eq.~\eqref{S:J}, which we do below.

%%%%%%%%%%%%%%%%%%%%%%%%%%%%%%%%%%%%%%%%%%%%%%%%%%%%%%%%%%%%%%%%%%%%%%%%%%%%%%%

\section{Equilibrium Wigner function and momentum integrals}
\label{sec:S6}

All momentum integrals are taken over the positive-energy branch, with the measure $\int_P\equiv2\int d^4p\,\theta(p^0)/(2\pi)^3$ of Sec.~\ref{sec:S4}. Antiparticles enter through their own current $\bar J^\mu_\chi$, Eq.~\eqref{S:anti}, so the negative-energy branch and its vacuum term are never needed.

\paragraph{Equilibrium currents.}
Inserting Eq.~\eqref{S:fcomoving} into Eq.~\eqref{S:J} with $n=u$, and using $\nabla_\nu f^{(0)}_\chi=f^{(0)}_\chi(1-f^{(0)}_\chi)\,\vp_{\nu\lambda}p^\lambda$, gives the particle current
\begin{equation}
J^\mu_\chi=\delta(p^2)\Big[p^\mu f^{(0)}_\chi-\frac{\chi\hbar}{2T}\frac{\omega\cdot p}{u\cdot p}\,p^\mu f^{(0)}_\chi\big(1-f^{(0)}_\chi\big)+\chi\hbar\,S^{\mu\nu}_{(u)}\vp_{\nu\lambda}p^\lambda f^{(0)}_\chi\big(1-f^{(0)}_\chi\big)\Big]+\chi\hbar Q\,\Ft^{\mu\nu}p_\nu\,\delta'(p^2)\,f^{(0)}_\chi .
\label{S:Jeq}
\end{equation}
The three $\mathcal O(\hbar)$ terms are the spin-vorticity correction to the distribution, the side-jump current, and the magnetic-moment term. For antiparticles, the solution $\bar g_\chi$ of Sec.~\ref{sec:S5} gives $\bar f_\chi=\bar f^{(0)}_\chi+\frac{\chi\hbar}{2T}\frac{\omega\cdot p}{u\cdot p}\,\bar f^{(0)}_\chi(1-\bar f^{(0)}_\chi)$, with $\bar f^{(0)}_\chi=[e^{(u\cdot p+\mu_\chi)/T}+1]^{-1}$. In $\bar J^\mu_\chi$ the derivative is $\bar\nabla_\mu=\partial_\mu+QF_{\mu\lambda}\partial_p^\lambda$. Because $\partial_\mu\alpha_\chi=-QF_{\mu\lambda}\beta^\lambda$, the field terms cancel, and $\bar\nabla_\nu\bar f^{(0)}_\chi=\bar f^{(0)}_\chi(1-\bar f^{(0)}_\chi)\,\vp_{\nu\lambda}p^\lambda$. Hence
\begin{equation}
\bar J^\mu_\chi=\delta(p^2)\Big[p^\mu\bar f^{(0)}_\chi+\frac{\chi\hbar}{2T}\frac{\omega\cdot p}{u\cdot p}\,p^\mu\bar f^{(0)}_\chi\big(1-\bar f^{(0)}_\chi\big)-\chi\hbar\,S^{\mu\nu}_{(u)}\vp_{\nu\lambda}p^\lambda\bar f^{(0)}_\chi\big(1-\bar f^{(0)}_\chi\big)\Big]+\chi\hbar Q\,\Ft^{\mu\nu}p_\nu\,\delta'(p^2)\,\bar f^{(0)}_\chi ,
\label{S:Jbareq}
\end{equation}
which is Eq.~\eqref{S:Jeq} with $(\mu_\chi,Q,\chi)\to(-\mu_\chi,-Q,-\chi)$. The two vortical terms change sign because the antiparticle of a chirality-$\chi$ fermion has helicity $-\chi/2$. The magnetic-moment term, proportional to $\chi Q$, keeps its sign. The moments below are evaluated at a point where $E^\mu=F^{\mu\nu}u_\nu=0$ (ideal magnetohydrodynamics, or the rotation axis). There $\Ft^{\mu\nu}p_\nu=B^\mu(u\cdot p)-u^\mu(B\cdot p)$.

\paragraph{Moments.}
Both $J^\mu_\chi$ and $\bar J^\mu_\chi$ are number currents. A macroscopic moment adds the antiparticle term with the sign of the quantity that an antiparticle carries:
\begin{equation}
j^\mu_\chi=\int_P\big(J^\mu_\chi-\bar J^\mu_\chi\big),\qquad
T^{\mu\nu}=\sum_\chi\int_Pp^\nu\big(J^\mu_\chi+\bar J^\mu_\chi\big).
\label{S:moments}
\end{equation}
The charge current $Qj^\mu_\chi$ counts particles minus antiparticles, while $T^{\mu\nu}$, like the entropy current~\eqref{S:Smu}, counts their sum. The axial current is $j^\mu_5=\sum_\chi\chi j^\mu_\chi=\int_P(A^\mu-\bar A^\mu)$, with $A^\mu=\sum_\chi\chi J^\mu_\chi$ and $\bar A^\mu=\sum_\chi\chi\bar J^\mu_\chi$, and $s^\mu=\frac\hbar2j^\mu_5$ is the spin density (Sec.~\ref{sec:S9}). Let $X^+(\mu_\chi,Q,\chi)=\int_Pw(p)\,J^\mu_\chi$ be the particle part of a moment of one chirality. By Eq.~\eqref{S:Jbareq}, the full moment is
\begin{equation}
X=X^{+}(\mu_\chi,Q,\chi)+\eta\,X^{+}(-\mu_\chi,-Q,-\chi),
\label{S:antirule}
\end{equation}
with $\eta=-1$ for $j^\mu_\chi$ and $\eta=+1$ for $T^{\mu\nu}$. Integrating instead over both energy branches with $\int_P\equiv2\int d^4p/(2\pi)^3$, and dropping the vacuum term, gives the same result. By Eq.~\eqref{S:anti}, the $p^0<0$ part of $\int_Pw(p)\,J^\mu_\chi$, for a weight $w$ homogeneous of degree $k$ in $p$, equals $(-1)^{k+1}\int_Pw(p)\,\bar J^\mu_\chi$. This reproduces $\eta=-1$ for the currents ($k=0$) and $\eta=+1$ for $T^{\mu\nu}$ ($k=1$).

\paragraph{Integration rules.}
For $|\bm p|>0$ the step function equals unity in a neighborhood of the support of $\delta(p^2)$ and $\delta'(p^2)$ on the positive branch. Hence
\begin{equation}
\int_P\delta(p^2)\,G(p)=\int\!\frac{d^3p}{(2\pi)^3|\bm p|}\,G\big|_{p^0=|\bm p|},\qquad
\int_P\delta'(p^2)\,C(p)=-\int\!\frac{d^3p}{(2\pi)^3|\bm p|}\,\frac{\partial}{\partial p^0}\Big[\frac{C(p)}{2p^0}\Big]_{p^0=|\bm p|}.
\label{S:p0rules}
\end{equation}
The second relation follows from $\delta'(p^2)=(2p^0)^{-1}\partial_{p^0}\delta(p^2)$ and an integration by parts in $p^0$; the derivative $\partial_{p^0}$ also acts on the $u\cdot p$ dependence of $f^{(0)}_\chi$ and $\bar f^{(0)}_\chi$. The derivative of $\theta(p^0)$ contributes only at the tip of the cone, $p^\mu=0$, which has measure zero for every moment considered here. The tip matters only for total momentum derivatives, which yield the anomaly [property (c) of Sec.~\ref{sec:S3}]. Angular integrals use $\langle\hat p^i\hat p^j\rangle=\delta^{ij}/3$ and $\langle\hat p^i\hat p^j\hat p^k\hat p^l\rangle=(\delta^{ij}\delta^{kl}+\delta^{ik}\delta^{jl}+\delta^{il}\delta^{jk})/15$. Odd moments vanish. The radial integrals are Fermi integrals,
\begin{equation}
I_n(\mu)\equiv\int_0^\infty dk\,\frac{k^n}{e^{(k-\mu)/T}+1}=-\,n!\,T^{n+1}{\rm Li}_{n+1}\big(-e^{\mu/T}\big),\qquad
\int_0^\infty dk\,k^nf^{(0)}(1-f^{(0)})=nT\,I_{n-1}(\mu)\quad(n\geq1),
\label{S:Fermi}
\end{equation}
where the second relation follows from $f^{(0)}(1-f^{(0)})=-T\partial_kf^{(0)}$. Antiparticle integrals are the same functions evaluated at $-\mu_\chi$. Combined according to Eq.~\eqref{S:antirule}, they always appear as $I_n(\mu)+(-1)^{n+1}I_n(-\mu)$. By the inversion formula ${\rm Li}_s(-e^{x})+(-1)^s{\rm Li}_s(-e^{-x})=-\frac{(2\pi i)^s}{s!}B_s\big(\tfrac12+\tfrac{x}{2\pi i}\big)$, where $B_s$ are Bernoulli polynomials, these combinations are polynomials:
\begin{align}
I_0(\mu)-I_0(-\mu)&=\mu, &
I_1(\mu)+I_1(-\mu)&=\frac{\mu^2}{2}+\frac{\pi^2T^2}{6},\nonumber\\
I_2(\mu)-I_2(-\mu)&=\frac{\mu^3}{3}+\frac{\pi^2T^2\mu}{3}, &
I_3(\mu)+I_3(-\mu)&=\frac{\mu^4}{4}+\frac{\pi^2T^2\mu^2}{2}+\frac{7\pi^4T^4}{60}.
\label{S:polylog}
\end{align}
All results below are therefore exact in $\mu_\chi/T$; no degenerate or high-temperature expansion is involved. 

%%%%%%%%%%%%%%%%%%%%%%%%%%%%%%%%%%%%%%%%%%%%%%%%%%%%%%%%%%%%%%%%%%%%%%%%%%%%%%%
\section{Vector and axial currents}
\label{sec:S7}

We evaluate $j^\mu_\chi$, Eq.~\eqref{S:moments}, with Eqs.~\eqref{S:Jeq} and \eqref{S:Jbareq} in the local rest frame, $u^\mu=(1,\bm 0)$, where $S^{ij}_{(u)}=\epsilon^{ijk}p^k/(2|\bm p|)$, $S^{0i}_{(u)}=0$, $\vp_{ij}=\epsilon^{ijk}\omega^k/T$, and $\Ft^{i\nu}p_\nu=B^i p^0$. The particle contributions are as follows.

\noindent(i) Ideal part:
\begin{equation}
\int_P\delta(p^2)\,p^\mu f^{(0)}_\chi=\frac{u^\mu}{2\pi^2}\,I_2(\mu_\chi).
\end{equation}
(ii) Spin-vorticity term. Since $-\omega\cdot p/(u\cdot p)=\bm\omega\cdot\hat{\bm p}$,
\begin{equation}
\frac{\chi\hbar}{2T}\int\!\frac{d^3p}{(2\pi)^3|\bm p|}\,p^i(\bm\omega\cdot\hat{\bm p})f^{(0)}_\chi(1-f^{(0)}_\chi)=\frac{\chi\hbar\,\omega^i}{6\pi^2}\,I_1(\mu_\chi).
\end{equation}
(iii) Side-jump term. Using $S^{ij}_{(u)}\vp_{jk}p^k=[|\bm p|^2\omega^i-p^i(\bm p\cdot\bm\omega)]/(2|\bm p|T)$, whose angular average is $|\bm p|\omega^i/(3T)$, and noting that the $\vp_{j0}$ terms are odd in $\hat{\bm p}$,
\begin{equation}
\chi\hbar\int_P\delta(p^2)\,S^{i\nu}_{(u)}\vp_{\nu\lambda}p^\lambda f^{(0)}_\chi(1-f^{(0)}_\chi)=\frac{\chi\hbar\,\omega^i}{3\pi^2}\,I_1(\mu_\chi).
\end{equation}
(iv) Magnetic-moment term. By Eq.~\eqref{S:p0rules},
\begin{equation}
\chi\hbar QB^i\int_Pp^0\delta'(p^2)f^{(0)}_\chi=-\frac{\chi\hbar QB^i}{4\pi^2}\int_0^\infty dk\,k\,\partial_kf^{(0)}_\chi=\frac{\chi\hbar QB^i}{4\pi^2}\,I_0(\mu_\chi).
\end{equation}
The time component of (iv) is odd in $\hat{\bm p}$ and vanishes.

\noindent The vortical response thus comes one third from the distribution function and two thirds from the side jump, and the magnetic response comes entirely from the magnetic-moment term. Adding antiparticles with Eq.~\eqref{S:antirule} ($k=0$) and using Eq.~\eqref{S:polylog},
\begin{equation}
j^\mu_\chi=n_\chi u^\mu+\chi\hbar\Big(\frac{\mu_\chi^2}{4\pi^2}+\frac{T^2}{12}\Big)\omega^\mu+\chi\hbar\,\frac{Q\mu_\chi}{4\pi^2}B^\mu,\qquad
n_\chi=\frac{\mu_\chi^3}{6\pi^2}+\frac{\mu_\chi T^2}{6}.
\label{S:jchi}
\end{equation}
These are the CVE and CME of a single Weyl fermion in the thermodynamic frame defined by $\beta^\mu$~\cite{Vilenkin:1979ui,Vilenkin:1980fu,Landsteiner:2011cp,Stephanov:2015roa}. The vector and axial currents, $j^\mu=\sum_\chi j^\mu_\chi$ and $j^\mu_5=\sum_\chi\chi j^\mu_\chi$ with $\mu_\pm=\mu\pm\mu_5$, follow from $\sum_\chi\mu_\chi=2\mu$, $\sum_\chi\chi\mu_\chi=2\mu_5$, $\sum_\chi\mu_\chi^2=2(\mu^2+\mu_5^2)$, $\sum_\chi\chi\mu_\chi^2=4\mu\mu_5$, $\sum_\chi\mu_\chi^3=2\mu^3+6\mu\mu_5^2$, and $\sum_\chi\chi\mu_\chi^3=6\mu^2\mu_5+2\mu_5^3$. The result is Eqs.~\eqref{eq:currents1}, \eqref{eq:currents} and \eqref{eq:coeffs}.

\paragraph{Relations among the coefficients.}
For each chirality, Eq.~\eqref{S:jchi} gives
\begin{equation}
\chi\Big(\frac{\mu_\chi^2}{4\pi^2}+\frac{T^2}{12}\Big)=\frac{\chi}{2}\frac{\partial n_\chi}{\partial\mu_\chi},\qquad
\frac{Q\mu_\chi}{4\pi^2}=\frac Q2\frac{\partial}{\partial\mu_\chi}\Big(\frac{\mu_\chi^2}{4\pi^2}+\frac{T^2}{12}\Big).
\label{S:relchi}
\end{equation}
Summing over $\chi$ with $\partial/\partial\mu=\sum_\chi\partial/\partial\mu_\chi$ yields $\xi=\frac12\partial n_5/\partial\mu$, $\xi_5=\frac12\partial n/\partial\mu$, $\xi_B=\frac Q2\partial\xi/\partial\mu$, and $\xi_{B5}=\frac Q2\partial\xi_5/\partial\mu$. The first relation reflects the fact that the spin-vorticity coupling $-\chi\hbar\,\hat{\bm p}\cdot\bm\omega/2$ acts as a helicity-dependent shift of the chemical potential.

%%%%%%%%%%%%%%%%%%%%%%%%%%%%%%%%%%%%%%%%%%%%%%%%%%%%%%%%%%%%%%%%%%%%%%%%%%%%%%%
\section{Energy-momentum tensor, spin tensor, and entropy current}
\label{sec:S8}

\paragraph{Energy-momentum tensor.}
The tensor $T^{\mu\nu}$ of Eq.~\eqref{S:moments} is the canonical (kinetic) tensor of the Dirac field, whose first index labels the flux. Proceeding as in Sec.~\ref{sec:S7}, now with the four-index angular average and Eq.~\eqref{S:antirule} with $\eta=+1$, the contributions of each chirality are as follows.

\noindent(i) Ideal part:
\begin{equation}
\varepsilon_\chi u^\mu u^\nu-P_\chi\Delta^{\mu\nu},\qquad
\varepsilon_\chi=3P_\chi=\frac{I_3(\mu_\chi)+I_3(-\mu_\chi)}{2\pi^2}=\frac{7\pi^2T^4}{120}+\frac{\mu_\chi^2T^2}{4}+\frac{\mu_\chi^4}{8\pi^2}.
\end{equation}
(ii) Spin-vorticity term: $\frac{\chi\hbar n_\chi}{2}\big(\omega^\mu u^\nu+u^\mu\omega^\nu\big)$.\\
(iii) Side-jump term: $\chi\hbar n_\chi\big(\omega^\mu u^\nu-\frac12\epsilon^{\mu\nu\alpha\beta}u_\alpha a_\beta\big)$.\\
(iv) Magnetic-moment term: $\chi\hbar Q\big(\frac{\mu_\chi^2}{8\pi^2}+\frac{T^2}{24}\big)\big(B^\mu u^\nu+u^\mu B^\nu\big)$.

\noindent For instance, the $(i,0)$ component of (iv) is $\chi\hbar QB^i\int_P(p^0)^2\delta'(p^2)f^{(0)}_\chi=\chi\hbar QB^iI_1(\mu_\chi)/(4\pi^2)$, and the $(0,i)$ component gives the same value. In (iii), the term with $a_\beta$ arises from the $\vp_{j0}$ components of the thermal vorticity contribution, $\frac{\chi\hbar n_\chi}{2}\epsilon^{ijk}a^k$, to $T^{ij}$. Summing over chiralities gives Eqs.~\eqref{eq:Tmunu} and \eqref{eq:eps}, with
\begin{equation}
\varepsilon=\sum_\chi\varepsilon_\chi=\frac{7\pi^2T^4}{60}+\frac{T^2(\mu^2+\mu_5^2)}{2}+\frac{\mu^4+6\mu^2\mu_5^2+\mu_5^4}{4\pi^2},\qquad
q^\mu=n_5\,\omega^\mu+\frac{Q\mu\mu_5}{2\pi^2}B^\mu .
\end{equation}
The coefficient $7\pi^2/60$ is that of a massless Dirac fermion, i.e., two Weyl species including antiparticles. The symmetric part of $T^{\mu\nu}$ carries the energy flow $\hbar q^\mu$ found in field theory~\cite{Landsteiner:2011cp}, while its antisymmetric part is
\begin{equation}
T^{\mu\nu}-T^{\nu\mu}=\hbar n_5\big(\omega^\mu u^\nu-u^\mu\omega^\nu\big)+\hbar n_5\,\epsilon^{\mu\nu\rho\sigma}a_\rho u_\sigma .
\label{S:Tanti}
\end{equation}

\paragraph{Spin tensor and angular-momentum conservation.}
For massless fermions the spin density is $\bm s=\hbar\bm j_5/2$. The canonical spin tensor with $S^{0ij}=\epsilon^{ijk}s^k$ is $S^{\lambda\mu\nu}=-\frac\hbar2\epsilon^{\lambda\mu\nu\sigma}j_{5\sigma}$, Eq.~\eqref{eq:spin}. To leading order, $j_5^\sigma=n_5u^\sigma$. In global equilibrium $\mu_\chi/T$ is constant and $\partial_\lambda T=Ta_\lambda$, so the homogeneous function $n_5$ of degree three obeys $\partial_\lambda n_5=3n_5a_\lambda$. Together with $\partial_\lambda u_\sigma=a_\lambda u_\sigma-T\vp_{\lambda\sigma}$ and $\frac12\epsilon^{\mu\nu\lambda\sigma}\vp_{\lambda\sigma}=-(\omega^\mu u^\nu-u^\mu\omega^\nu)/T+\epsilon^{\mu\nu\rho\sigma}a_\rho u_\sigma/T$, this gives
\begin{equation}
\partial_\lambda S^{\lambda\mu\nu}=-\frac\hbar2\,\partial_\lambda\big(\epsilon^{\lambda\mu\nu\sigma}n_5u_\sigma\big)
=-\hbar n_5\big(\omega^\mu u^\nu-u^\mu\omega^\nu\big)-\hbar n_5\,\epsilon^{\mu\nu\rho\sigma}a_\rho u_\sigma
=T^{\nu\mu}-T^{\mu\nu},
\label{S:AM}
\end{equation}
in agreement with Eq.~\eqref{S:Tanti}. Total angular momentum $J^{\lambda\mu\nu}=x^\mu T^{\lambda\nu}-x^\nu T^{\lambda\mu}+S^{\lambda\mu\nu}$ is therefore conserved. This is a nontrivial check of the side-jump and frame structure of the first-order solution. Pseudogauge transformations~\cite{Li:2020eon,Florkowski:2018fap} can be used to symmetrize $T^{\mu\nu}$; the Belinfante tensor is $\varepsilon u^\mu u^\nu-P\Delta^{\mu\nu}+\hbar(q^\mu u^\nu+u^\mu q^\nu)$.

\paragraph{Entropy current.}
Evaluating Eq.~\eqref{S:Smu} with the rules of Sec.~\ref{sec:S6} (now with $\Phi$ in place of $f$, so that $\Phi'(f^{(0)})\nabla_\nu f^{(0)}$ and $\Phi(f^{(0)})$ enter), we find for each chirality
\begin{equation}
\mathcal H^\mu_\chi=s_\chi u^\mu+\frac{\chi\hbar T}{12}\big(2\mu_\chi\,\omega^\mu+Q B^\mu\big),\qquad
s_\chi=\frac{\varepsilon_\chi+P_\chi-\mu_\chi n_\chi}{T}=\frac{7\pi^2T^3}{90}+\frac{\mu_\chi^2T}{6}.
\end{equation}
The magnetic terms cancel between the two chiralities, and the sum is Eq.~\eqref{eq:entropy}. Its divergence vanishes explicitly: $\partial_\mu(su^\mu)=0$ because $u\cdot\partial T=0$ and $\partial\cdot u=0$, and $\partial_\mu(T\mu_5\omega^\mu)=2T\mu_5\,a\cdot\omega+T\mu_5\,\partial\cdot\omega=0$ because $\partial\cdot\omega=-2a\cdot\omega$ for a Killing flow. This confirms Eq.~\eqref{S:dS} at the level of moments.

%%%%%%%%%%%%%%%%%%%%%%%%%%%%%%%%%%%%%%%%%%%%%%%%%%%%%%%%%%%%%%%%%%%%%%%%%%%%%%%
\section{Spin polarization}
\label{sec:S9}

\paragraph{Axial Wigner function.}
With the spin tensor~\eqref{eq:spin}, the spin density is $s^\mu=\frac\hbar2\int_P(A^\mu-\bar A^\mu)$, where $A^\mu=\sum_\chi\chi J^\mu_\chi$ and $\bar A^\mu=\sum_\chi\chi\bar J^\mu_\chi$, Eq.~\eqref{S:moments}. Hence $\frac\hbar2A^\mu$ and $-\frac\hbar2\bar A^\mu$ are the phase-space densities of the spin of particles and antiparticles. For an arbitrary frame vector, the equilibrium distribution~\eqref{S:geq} contains the first-order correction $\frac{\chi\hbar}{2}(\vp_{\alpha\beta}S^{\alpha\beta}_{(n)})f^{(0)}_\chi(1-f^{(0)}_\chi)$, while the corresponding side-jump contribution is $\chi\hbar S^{\mu\nu}_{(n)}\nabla_\nu f^{(0)}_\chi=\chi\hbar S^{\mu\nu}_{(n)}\vp_{\nu\lambda}p^\lambda f^{(0)}_\chi(1-f^{(0)}_\chi)$. The first-order vortical part of $J^\mu_\chi$ is therefore $\chi\hbar\,\delta(p^2)f^{(0)}_\chi(1-f^{(0)}_\chi)X^\mu$, with
\begin{equation}
X^\mu\equiv S^{\mu\nu}_{(n)}\vp_{\nu\lambda}p^\lambda+\frac12p^\mu\vp_{\alpha\beta}S^{\alpha\beta}_{(n)}=-\frac14\,\epsilon^{\mu\nu\rho\sigma}p_\nu\vp_{\rho\sigma}\qquad(p^2=0).
\label{S:X}
\end{equation}
By Eq.~\eqref{S:eqframe}, $X^\mu$ is independent of $n$. The covariant form then follows from the Schouten identity. Since $\chi^2=1$, the first-order terms of both chiralities add, and the particle branch gives
\begin{equation}
A^\mu=p^\mu\delta(p^2)\big(f^{(0)}_+-f^{(0)}_-\big)-\frac\hbar4\,\epsilon^{\mu\nu\rho\sigma}p_\nu\vp_{\rho\sigma}\,\delta(p^2)\sum_\chi f^{(0)}_\chi(1-f^{(0)}_\chi)+\hbar Q\,\Ft^{\mu\nu}p_\nu\,\delta'(p^2)\sum_\chi f^{(0)}_\chi ,
\label{S:Aeq}
\end{equation}
which is Eq.~\eqref{eq:Aeq}. Every term is independent of the frame vector: the frame dependence of the distribution and that of the side-jump current cancel, as they must for an observable. The vortical term is the $m\to0$ limit of the global-equilibrium result for massive Dirac fermions, $A^\mu=-\frac\hbar4\epsilon^{\mu\nu\rho\sigma}p_\nu\vp_{\rho\sigma}f(1-f)\delta(p^2-m^2)$~\cite{Becattini:2013fla,Fang:2016vpj,Weickgenannt:2019dks}.

\paragraph{Rest-frame polarization.}
With Eq.~\eqref{S:vpdecomp}, $\epsilon^{i\nu\rho\sigma}p_\nu\vp_{\rho\sigma}=-\frac2T(p^0\bm\omega+\bm p\times\bm a)^i$, and $\Ft^{i\nu}p_\nu=B^ip^0$ for $E^\mu=0$. Integrating over $p^0$ with Eq.~\eqref{S:p0rules}, the particle density per unit $d^3p/(2\pi)^3$ is $\sum_\chi f_\chi$. The spin-polarization density per unit $d^3p/(2\pi)^3$ is
\begin{equation}
\bm{\mathcal A}(\bm p)=\hat{\bm p}\big(f_+-f_-\big)+\frac{\hbar}{2T}\Big(\bm\omega+\hat{\bm p}\times\bm a+\frac{Q\bm B}{|\bm p|}\Big)\sum_\chi f_\chi(1-f_\chi),
\label{S:Arest}
\end{equation}
where $f_\chi\equiv f^{(0)}_\chi(|\bm p|)$ and the magnetic term follows from $-\frac{1}{2|\bm p|}\partial_{p^0}[p^0f/(2p^0)]=f(1-f)/(4|\bm p|T)$. The mean polarization is $\bm P=\bm{\mathcal A}/\sum_\chi f_\chi$, which is Eq.~\eqref{eq:Prest}. By Eq.~\eqref{S:anti}, antiparticles obey the same expression with $Q\to-Q$, $f_\chi\to\bar f_\chi$, and $f_+-f_-\to\bar f_--\bar f_+$ (the antiparticle of a right-handed fermion has negative helicity). As a check, integrating $\frac\hbar2\bm{\mathcal A}$ over momentum with antiparticles reproduces the spin density $\frac\hbar2\bm j_5=\frac{\hbar^2}{2}(\xi_5\bm\omega+\xi_{B5}\bm B)$. The vortical term uses $I_1(\mu)+I_1(-\mu)$; the magnetic term uses $I_0(\mu)-I_0(-\mu)$, because the field polarizes antiparticles oppositely; the $\hat{\bm p}\times\bm a$ term averages to zero.

\paragraph{Phenomenology.}
For nondegenerate distributions, $\langle1-f\rangle\simeq1$. The vortical part $\frac{\hbar}{2T}\bm\omega$ is then common to particles and antiparticles and reduces to $\frac12\bm\varpi$ in the usual notation, as for massive particles. The acceleration term $\hat{\bm p}\times\bm a$ produces a momentum-dependent polarization pattern. The magnetic term splits particles and antiparticles, $\bm P-\bar{\bm P}\simeq\hbar Q\bm B/(|\bm p|T)$. This is the chiral-fermion analogue of the $\Lambda$--$\bar\Lambda$ splitting $2\mu_\Lambda\bm B/T$ proposed as a magnetometer in heavy-ion collisions~\cite{Becattini:2016gvu}, with the moment $\mu_\Lambda$ replaced by $\hbar Q/(2|\bm p|)$, so that soft quarks are polarized most strongly. In Weyl semimetals, Eq.~\eqref{S:Arest} gives the pseudospin polarization of the Weyl quasiparticles induced by rotation and by a magnetic field.

%%%%%%%%%%%%%%%%%%%%%%%%%%%%%%%%%%%%%%%%%%%%%%%%%%%%%%%%%%%%%%%%%%%%%%%%%%%%%%%
\section{Magnetovortical thermodynamics and magnetization}
\label{sec:S10}

\paragraph{Thermodynamic potential.}
The distribution of Sec.~\ref{sec:S5} is, to first order in $\hbar$, the Wigner function of the global-equilibrium density operator $\hat\rho=Z^{-1}\exp[-b\cdot\hat P+\frac12\vp_{\mu\nu}\hat J^{\mu\nu}+\sum_\chi\alpha_\chi\hat Q_\chi]$~\cite{Becattini:2012tc,Becattini:2013fla}. On the rotation axis and in the rest frame, $T\ln Z=PV$ with
\begin{equation}
dP=s\,dT+\sum_\chi n_\chi d\mu_\chi+\bm J\cdot d\bm\omega+\bm M\cdot d\bm B ,
\label{S:dP}
\end{equation}
where $\bm J$ is the angular-momentum density conjugate to $\bm\omega$ and $\bm M$ is the magnetization. In the comoving frame, the orbital angular momentum is carried by the flow $u^\mu$, and the local angular-momentum density is the spin density, $J^i=\frac12\epsilon^{ijk}S^{0jk}=\frac\hbar2j_5^i$. By Eq.~\eqref{S:jchi},
\begin{equation}
\bm J=\frac{\hbar^2}{2}\big(\xi_5\,\bm\omega+\xi_{B5}\bm B\big)+\mathcal O(\hbar^3).
\label{S:Jang}
\end{equation}
Since $\partial J_i/\partial\omega_j$ is symmetric, Eq.~\eqref{S:Jang} is integrable. The Maxwell relation $\partial M_j/\partial\omega_i=\partial J_i/\partial B_j$ then fixes every term of $P$ bilinear in $\bm\omega$ and $\bm B$, except the susceptibility $\chi_B$:
\begin{equation}
P=P_0(T,\mu_\chi)+\frac{\hbar^2}{4}\,\xi_5\,\bm\omega^2+\frac{\hbar^2}{2}\,\xi_{B5}\,\bm\omega\cdot\bm B+\frac12\chi_B\bm B^2+\cdots ,
\label{S:P2}
\end{equation}
which is Eq.~\eqref{eq:P2}.

\paragraph{Second-order responses.}
Differentiating Eq.~\eqref{S:P2} with $\xi_5=(\mu^2+\mu_5^2)/(2\pi^2)+T^2/6$ and $\xi_{B5}=Q\mu/(2\pi^2)$ gives
\begin{align}
\bm M&=\frac{\hbar^2}{2}\xi_{B5}\,\bm\omega+\chi_B\bm B=\frac{\hbar^2Q\mu}{4\pi^2}\,\bm\omega+\chi_B\bm B, \label{S:M}\\
\delta n&=\frac{\hbar^2\mu}{4\pi^2}\,\bm\omega^2+\frac{\hbar^2Q}{4\pi^2}\,\bm\omega\cdot\bm B+\frac12\frac{\partial\chi_B}{\partial\mu}\bm B^2,\qquad
\delta n_5=\frac{\hbar^2\mu_5}{4\pi^2}\,\bm\omega^2+\frac12\frac{\partial\chi_B}{\partial\mu_5}\bm B^2, \label{S:dn}\\
\delta s&=\frac{\hbar^2T}{12}\,\bm\omega^2+\frac12\frac{\partial\chi_B}{\partial T}\bm B^2,\qquad
\delta\varepsilon\big|_{\bm B=0}=T\delta s+\mu\,\delta n+\mu_5\delta n_5+\bm\omega\cdot\bm J-\delta P=\frac{3\hbar^2}{4}\,\xi_5\,\bm\omega^2 . \label{S:deps}
\end{align}
The last equality uses the Euler relation for the degree-two homogeneous function $\xi_5$. In covariant form, Eq.~\eqref{S:M} reads, for $E^\mu=0$,
\begin{equation}
M^{\mu\nu}=-\frac{\hbar^2T}{2}\,\xi_{B5}\,\Delta^\mu_{\ \alpha}\Delta^\nu_{\ \beta}\,\vp^{\alpha\beta}+\chi_BF^{\mu\nu}
=-\frac{\hbar^2Q\mu T}{4\pi^2}\Big[\vp^{\mu\nu}-\big(e^\mu u^\nu-e^\nu u^\mu\big)\Big]+\chi_BF^{\mu\nu},
\label{S:Mcov}
\end{equation}
which is Eq.~\eqref{eq:M}. Here $e^\mu\equiv\vp^{\mu\nu}u_\nu=a^\mu/T$ is the electric (acceleration) part of the thermal vorticity, which the projectors remove. From Eq.~\eqref{S:vpdecomp}, $\Delta^\mu_{\ \alpha}\Delta^\nu_{\ \beta}\vp^{\alpha\beta}=-\epsilon^{\mu\nu\alpha\beta}u_\alpha\omega_\beta/T$. With $M^{ij}=-\epsilon^{ijk}M^k$ and $\vp^{ij}=\epsilon^{ijk}\omega^k/T$ in the rest frame, Eq.~\eqref{S:Mcov} gives $M^{ij}=-\frac{\hbar^2T}{2}\xi_{B5}\vp^{ij}$, i.e., Eq.~\eqref{S:M}. The rotation-induced magnetization is purely magnetic in the comoving frame, $M^{\mu\nu}u_\nu=0$.

\paragraph{Checks.}
(i) For $\mu_5=0$, Eqs.~\eqref{S:dn} and \eqref{S:deps} give $\delta n=\hbar^2\mu\bm\omega^2/(4\pi^2)$ and $\delta\varepsilon=\hbar^2(T^2/8+3\mu^2/8\pi^2)\bm\omega^2$. These coincide with the exact second-order coefficients of the free massless Dirac field in global equilibrium~\cite{Buzzegoli:2017cqy}: $\rho=\rho_0-w^2U_w$ and $n=n_0-w^2N_w$, with $U_w=\frac{T^4}{8}(1+\frac{3\mu^2}{\pi^2T^2})$, $N_w=\frac{\mu T^2}{4\pi^2}$, and $w^2=-\bm\omega^2/T^2$. This confirms that the angular-momentum density conjugate to $\bm\omega$ is the spin density.
(ii) The charge density bilinear in $\bm\omega$ and $\bm B$ is $\delta n=\frac{\hbar^2}{2}QC_A\,\bm\omega\cdot\bm B$, with the anomaly coefficient $C_A=1/(2\pi^2)$. It is independent of $T$ and $\mu$ and coincides with the result of Hattori and Yin, obtained in the lowest Landau level~\cite{Hattori:2016njk}. Equation~\eqref{S:dn} shows that it holds for weak fields as well; it is a direct consequence of the CSE and the Maxwell relation $\partial n/\partial B=\partial M/\partial\mu$.
(iii) At $\bm B=0$ the ratio of the rotation-induced magnetization to the rotation-induced spin density is $\xi_{B5}/\xi_5$. For $T\to0$ and $\mu_5=0$ it tends to $Q/\mu$, the gyromagnetic ratio $Q/E$ of a Dirac particle with $g=2$ evaluated at the Fermi surface. At $\mu=0$ it vanishes, because particles and antiparticles compensate.

\paragraph{Kinetic interpretation.}
The shell~\eqref{S:Ep} defines the magnetic moment. With $p^0=|\bm p|+\chi\hbar QF_{\alpha\beta}S^{\alpha\beta}_{(u)}/(2|\bm p|)\equiv|\bm p|-\frac12m^{\mu\nu}_\chi F_{\mu\nu}$,
\begin{equation}
m^{\mu\nu}_\chi=-\chi\hbar Q\,\frac{S^{\mu\nu}_{(u)}}{u\cdot p},\qquad
m^{ij}_\chi=-\epsilon^{ijk}m^k_\chi,\qquad
\bm m_\chi=\chi\hbar Q\,\frac{\hat{\bm p}}{2|\bm p|},
\label{S:moment}
\end{equation}
parallel to the spin for $Q>0$, with the gyromagnetic ratio of a Dirac particle of energy $|\bm p|$. Antiparticles have opposite charge and opposite helicity and therefore the same $\bm m_\chi$ relative to $\hat{\bm p}$.

\noindent(a) Dipole density. The dipole-moment density of the polarized quasiparticles, $\bm M_{\rm dip}=\sum_\chi\int\frac{d^3p}{(2\pi)^3}\bm m_\chi(f_\chi+\bar f_\chi)$ is obtained from the vorticity-induced corrections $\left[\chi\hbar(\bm\omega\cdot\hat{\bm p})f^{(0)}_\chi(1-f^{(0)}_\chi)/(2T)\right]$ and $\left[-\chi\hbar(\bm\omega\cdot\hat{\bm p})\bar f^{(0)}_\chi(1-\bar f^{(0)}_\chi)/(2T)\right]$ to the particle and antiparticle distributions, respectively. We obtain
\begin{equation}
\bm M_{\rm dip}=\frac{\hbar^2Q\,\bm\omega}{24\pi^2}\sum_\chi\big[I_0(\mu_\chi)-I_0(-\mu_\chi)\big]=\frac{\hbar^2Q\mu}{12\pi^2}\,\bm\omega ,
\label{S:Mdip}
\end{equation}
one third of Eq.~\eqref{S:M}. In the covariant language, $\bm M_{\rm dip}$ is the total-derivative part $\partial_\nu M^{\nu\mu}$ of the side-jump current at this order.\\
(b) Berry-curvature density of states. The quasiparticle grand potential, Eq.~\eqref{S:Omega} below with the spin-vorticity shift included in the exponents, yields the semiclassical magnetization $\bm M=\sum\int[\sqrt\gamma\,\bm m_\chi f+TQ\bm\Omega_\chi\ln(1+e^{-g})]$~\cite{Xiao:2010}. Its second term contributes another $\hbar^2Q\mu\bm\omega/(12\pi^2)$.\\
(c) Anomalous velocity. The same grand potential gives as $-\partial\Omega/\partial\bm\omega$ the helicity density of the quasiparticles, which equals only $\frac23$ of the spin density $\frac\hbar2\bm j_5$ along $\bm B$. The remaining third comes from the anomalous velocity $Q(\tilde{\bm v}\cdot\bm\Omega_\chi)\bm B$ in $\sqrt\gamma\,\dot{\bm x}$, which enters the current but not the helicity. Through the Maxwell relation it supplies the last third of $\bm M$.\\
Thus $\bm M=\bm M_{\rm dip}+\bm M_{\rm Berry}+\bm M_{\rm av}$, with each term equal to $\hbar^2Q\mu\bm\omega/(12\pi^2)$.

\paragraph{Field-induced magnetization.}
The susceptibility $\chi_B$ is not fixed by first-order coefficients. To estimate it we set $\bm\omega=0$ and work in the rest frame with a constant $\bm B$ and $\bm E=0$. The equilibrium state is then $f_\chi=n_F\big((E_{\bm p}-\mu_\chi)/T\big)$ on the shell~\eqref{S:Ep}, with phase-space density $\sqrt\gamma f_\chi$, Eq.~\eqref{S:measure}. By Eq.~\eqref{S:anti}, antiparticles of chirality $\chi$ have the same $E_{\bm p}$, $\sqrt\gamma$, and $\bm m_\chi$. The quasiparticle grand potential is
\begin{equation}
\Phi_G=-T\sum_\chi\int\!\frac{d^3p}{(2\pi)^3}\,\sqrt\gamma\,\Big[\ln\big(1+e^{-(E_{\bm p}-\mu_\chi)/T}\big)+\ln\big(1+e^{-(E_{\bm p}+\mu_\chi)/T}\big)\Big],
\label{S:Omega}
\end{equation}
with $E_{\bm p}=|\bm p|-\bm m_\chi\cdot\bm B$ and $\sqrt\gamma=1+\bm m_\chi\cdot\bm B/|\bm p|$. To linear order in $\bm B$, $\bm M=-\partial\Phi_G/\partial\bm B=\chi_B\bm B$ with
\begin{equation}
\chi_B=\frac{\hbar^2Q^2}{12\pi^2}\sum_\chi\int_0^\infty\frac{dk}{k}\,\big(f_\chi+\bar f_\chi\big)
+\frac{\hbar^2Q^2}{24\pi^2}\sum_\chi\big[f_\chi(0)+\bar f_\chi(0)\big],
\label{S:chiB}
\end{equation}
where $f_\chi=[e^{(k-\mu_\chi)/T}+1]^{-1}$ and $\bar f_\chi=[e^{(k+\mu_\chi)/T}+1]^{-1}$.

The second term is the Zeeman polarization of the moments by their energy $-\bm m_\chi\cdot\bm B$. Since $\bm m_\chi^2\propto1/|\bm p|^2$, the integral $\int d^3p\,\bm m_\chi^2(-\partial f/\partial|\bm p|)$ reduces to its boundary value at $|\bm p|=0$, and
\begin{equation}
f_\chi(0)+\bar f_\chi(0)=\frac{1}{e^{-\mu_\chi/T}+1}+\frac{1}{e^{\mu_\chi/T}+1}=1
\label{S:sumrule}
\end{equation}
gives $\chi_B^{\rm Z}=\hbar^2Q^2/(12\pi^2)$, independent of $T$, $\mu$, and $\mu_5$. A Curie law, $\chi_B\propto1/T$, arises when the moment does not depend on momentum, as for massive spin-$\frac12$ particles~\cite{Bhadury:2022ulr, Buzzegoli:2022qrr}. For massless chiral fermions the factor $1/T$ from $\partial f/\partial E$ is compensated by $\langle\bm m_\chi^2\rangle\propto1/|\bm p|^2$. A term $\propto\mu/T$ would appear only if the antiparticle moment were taken antiparallel to that of particles. Equation~\eqref{S:sumrule} would then be replaced by $f_\chi(0)-\bar f_\chi(0)=\tanh(\mu_\chi/2T)$, which contradicts Eq.~\eqref{S:anti} and violates charge conjugation.

The first term in Eq.~\eqref{S:chiB} comes from the Berry-curvature measure and from the field dependence of $E_{\bm p}$ inside the logarithms. It diverges logarithmically at small $k$, where $f_\chi+\bar f_\chi\to1$. Since the semiclassical expansion requires $|\bm p|\gg\sqrt{\hbar|QB|}$, it gives $\chi_B\simeq(\hbar^2Q^2/12\pi^2)\ln(T^2/\hbar|QB|)+\mathcal O(\hbar^2Q^2)$. This term lies beyond the accuracy of first-order CKT: the second-order field correction to the quasiparticle energy and the field-induced positional shift contribute at the same order~\cite{Gao:2015gyn}, and the complete orbital susceptibility of a Weyl node is logarithmically divergent and even in the chemical potential~\cite{Koshino:2016hiz}. The robust conclusions are that $\chi_B$ is even in $\mu$ and $\mu_5$ and depends on temperature at most logarithmically, with no Curie law.

\paragraph{Symmetry.}
Under charge conjugation, $(\mu,\mu_5,F_{\mu\nu},M_{\mu\nu})\to(-\mu,\mu_5,-F_{\mu\nu},-M_{\mu\nu})$ with $\omega^\mu$ unchanged. The coefficient of $F^{\mu\nu}$ in $M^{\mu\nu}$ must therefore be even in $\mu$, and that of the vortical term odd in $\mu$. Parity ($\mu_5\to-\mu_5$, with $\bm\omega$, $\bm B$, and $\bm M$ axial) requires both to be even in $\mu_5$. Equations~\eqref{S:M} and \eqref{S:chiB} satisfy these constraints, whereas a contribution $\propto\mu F^{\mu\nu}/T$ is excluded. In $P$, the term $\bm\omega\cdot\bm B$ is C-odd and P-even, so its coefficient must be odd in $\mu$, as $\xi_{B5}=Q\mu/(2\pi^2)$ is.

\end{document}